\documentclass[twoside]{article}
\usepackage{graphicx}
\usepackage{framed}
\usepackage{natbib}
\usepackage{amsmath}
\usepackage{lipsum} 
\usepackage{floatrow}
\newfloatcommand{capbtabbox}{table}[][\FBwidth]

\usepackage[sc]{mathpazo} 
\usepackage[T1]{fontenc} 
\usepackage{microtype} 

\usepackage[a4paper,hmarginratio=1:1,top=32mm,columnsep=12pt,textheight=245mm,textwidth=185mm]{geometry} 
\usepackage{multicol} 
\usepackage{nonfloat} 
\usepackage{hyperref} 

\usepackage[hang, small,labelfont=bf,up,textfont=it,up]{caption} 
\usepackage{booktabs} 
\usepackage{float} 

\usepackage{paralist} 

\usepackage{abstract} 

\usepackage{titlesec} 
\renewcommand\thesection{\Roman{section}}
\titleformat{\section}[block]{\large\scshape\centering}{\thesection.}{1em}{} 

\usepackage{fancyhdr} 
\usepackage[dvipsnames*,svgnames]{xcolor}
\usepackage[framemethod=TikZ]{mdframed}
\mdfdefinestyle{NR}{skipabove=\topskip,skipbelow=\topskip,%
                    ,align=center,%
                    innerleftmargin=.25cm,linecolor=black,%
                    linewidth=2pt,backgroundcolor=Linen}
\newmdenv[style=NR]{NR}
\mdfdefinestyle{NR1}{skipabove=\topskip,skipbelow=\topskip,
                    ,align=center,%
                    innerleftmargin=.25cm,linecolor=black,
                    innertopmargin=1pt,innerbottommargin=1pt,%
                    linewidth=2pt,backgroundcolor=LightGrey}
\newmdenv[style=NR1]{NR1}
\mdfdefinestyle{NR2}{skipabove=\topskip,skipbelow=\topskip,%
                    ,align=center,roundcorner=10pt,
                    innerleftmargin=.25cm,linecolor=black,%
                    linewidth=1pt,backgroundcolor=LightBlue}
\newmdenv[style=NR2]{NR2}

\newcommand\myfigure[1]{%
\medskip\noindent\begin{minipage}{\columnwidth}
\centering%
#1%
\end{minipage}\medskip}

\newcommand{\simlt}{\lower.5ex\hbox{$\; \buildrel < \over \sim \;$}}
\newcommand{\simgt}{\lower.5ex\hbox{$\; \buildrel > \over \sim \;$}}
\newcommand{\etal}{{\it et al.}}
\newcommand{\chronos}{\texttt{\emph{Chronos}}}
\def\HII{\hbox{H~$\scriptstyle\rm II\ $}}
\def\HI{\hbox{H~$\scriptstyle\rm I\ $}}

\title{\vspace{-4mm}
\textbf{\fontsize{24pt}{10pt}\selectfont
{\texttt{\emph{Chronos}}}\footnote{Chronos, the Greek god personifying time is
the name we have given to this survey, as it is designed to 
understand the formation and evolution of galaxies across cosmic time}}\\\vspace{+3mm} 
\fontsize{18pt}{10pt}\selectfont
A NIR Spectroscopic Galaxy Survey:\\
\vspace{+1mm}
\fontsize{16pt}{10pt}\selectfont
{\bf From the formation of galaxies to the peak of activity}
\vspace{+1mm}}
\author{\large
\textsc{Ignacio Ferreras$^1$}\thanks{i.ferreras@ucl.ac.uk}
\textsc{, Ray Sharples$^2$}\\
\vspace{+2mm}\\
{\bf Science Core Team}\\
\textsc{James S. Dunlop$^3$, Anna Pasquali$^4$, Francesco La Barbera$^5$, 
Alexander Vazdekis$^6$,}\\
\textsc{Sadegh Khochfar$^3$, Mark Cropper$^1$, Andrea Cimatti$^7$, Michele Cirasuolo$^{3,8}$,}\\
\vspace{+2mm}\\
\textsc{Richard Bower$^2$, Jarle Brinchmann$^9$, Ben Burningham$^{10}$, Michele Cappellari$^{11}$,}\\
\textsc{St\'ephane Charlot$^{12}$, Christopher J. Conselice$^{13}$, Emanuele Daddi$^{14}$, Eva K. Grebel$^4$,}\\
\textsc{Rob Ivison$^{3,8}$, Matt J. Jarvis$^{11}$, Daisuke Kawata$^1$, Robert C. Kennicutt$^{15}$,}\\
\textsc{Tom Kitching$^1$, Ofer Lahav$^{16}$, Roberto Maiolino$^{15}$, Mathew J. Page$^1$,}\\
\textsc{Reynier F. Peletier$^{17}$, Andrew Pontzen$^{16}$, Joseph Silk$^{12}$, Volker Springel$^{4,18}$,}\\
\textsc{Mark Sullivan$^{11}$, Ignacio Trujillo$^6$, Gillian Wright$^{8}$}
}
\date{Submitted to ESA, May 24$^{\rm th}$, 2013\\
\url{http://www.chronos-mission.eu}}
\vspace{-5mm}

\begin{document}

\setcounter{page}{2}
\maketitle 

\thispagestyle{fancy} 

\begin{figure}[h]
\begin{center}
\includegraphics[width=45mm]{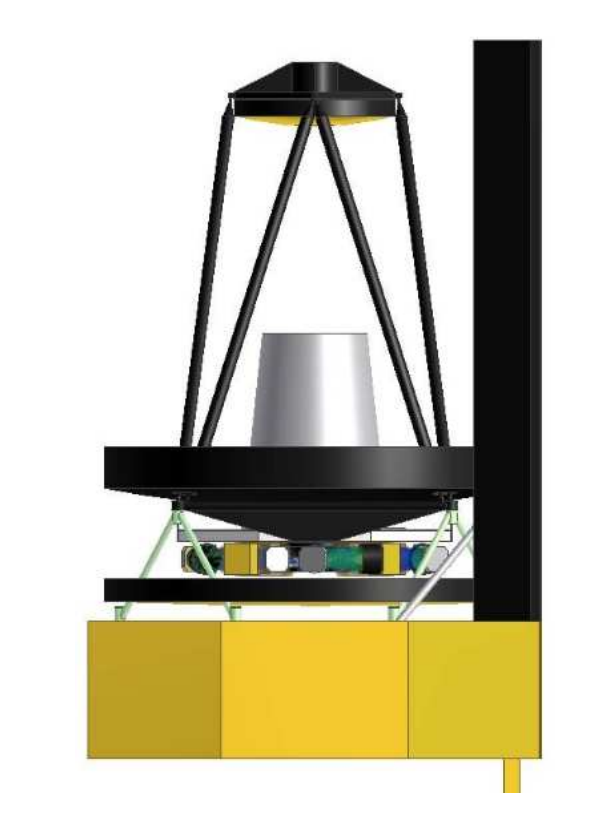}
\end{center}
\end{figure}

\noindent
{\small $^1$ Mullard Space Science Laboratory, University College London, UK; 
$^2$ University of Durham, UK;
$^3$ Royal Observatory, University of Edinburgh, UK;
$^4$ ARI/ZAH, University of Heidelberg, Germany;
$^5$ INAF/OAC, Naples, Italy;
$^6$ Instituto de Astrof\'\i sica de Canarias, Spain;
$^7$ Universit\'a di Bologna, Italy;
$^8$ UK ATC, Edinburgh, UK;
$^9$ Leiden University, the Netherlands;
$^{10}$ University of Hertfordshire, UK;
$^{11}$ Oxford University, UK;
$^{12}$ Institut d'Astrophysique de Paris, France;
$^{13}$ Nottingham University, UK;
$^{14}$ CEA/Saclay, France;
$^{15}$ Institute of Astronomy, University of Cambridge, UK;
$^{16}$ Physics \& Astronomy Department, University College London, UK;
$^{17}$ Kapteyn Institute, University of Groningen, the Netherlands;
$^{18}$ Heidelberg Institute for Theoretical Studies, Germany.
}

\clearpage


\begin{NR}[userdefinedwidth=164mm]
\begin{center}
{\bf\Large Executive Summary}
\end{center}

\vskip+2mm

In a decade or two from now, we will have made significant strides in
our understanding of the early formation history of the Universe,
through missions such as {\sl Planck} and {\sl Euclid}, and in its
recent state, through {\sl Gaia} and ground-based surveys such as SDSS
and their more substantial successors. What will still be problematic
is how we arrived here given the initial conditions. This is the realm
of ``baryon physics'', the nature of the formation and evolution of
the galaxies. Understanding this is a colossal task, currently
occupying a large fraction of the international astronomical
community. It involves complex astrophysics at redshifts 1--6. To make
the critical and substantial advances in our understanding of the
essential nature of this process requires spectroscopy (for the
astrophysics) in the infrared (because of the redshift) of a large
volume of the Universe (to examine the critical effects of
environment). This is what the concept presented here sets out to
achieve.
\medskip

We propose \chronos, an L-class mission to understand the formation and
evolution of galaxies, by collecting the deepest NIR spectroscopic data,
from the formation of the first galaxies at z$\sim$10 to the peak of
formation activity at z$\sim$1$-$3. The strong emission from 
the atmospheric background makes this type of survey impossible from
a ground-based observatory. The spectra of galaxies represent
the equivalent of a {\sl DNA fingerprint}, containing information
about the past history of star formation and chemical enrichment.
The proposed survey will allow us to dissect the 
formation process of galaxies including the timescales
of quenching triggered by star formation or AGN activity, 
the effect of environment, the role of infall/outflow processes, or
the connection between the galaxies and their underlying dark matter haloes.
To provide these data, the mission requires a 2.5m space
telescope optimised for a campaign of very deep NIR
spectroscopy. A combination
of a high multiplex and very long integration times will result
in the deepest, largest, high-quality spectroscopic dataset of
galaxies from z=$1$ to $12$, spanning the history of the Universe,
from 400\,million to 6\,billion years after the big bang, i.e. 
covering the most active half of cosmic history.
\medskip

The highly demanding requirements results in a mission that is the
spectroscopic equivalent of a {\sl Hubble Space Telescope} obtaining
one Ultra Deep Field (the deepest exposure of distant galaxies ever
attained) every fortnight for five years. A two-tiered survey will
provide a high quality stellar mass limited dataset of about {\sl 2
  million} spectra from galaxies covering the most important epochs of
structure formation, back to the early phases soon after
recombination. Although missions such as {\sl Euclid} or {\sl WFIRST}
will provide low-resolution spectra in the NIR, the requirements of
resolution and SNR in the continuum for the analysis of the properties
of the underlying stellar populations at z$\simlt$3 are too demanding
for them. Therefore, \chronos is the link between
cosmology-orientated missions, such as {\sl Planck} or {\sl Euclid}
and {\sl Gaia}'s targeted exploration of our own galaxy.  Our mission
will gather key spectroscopic indicators of the properties of the
underlying stellar populations in galaxies. This will be complemented
by {\sl Herschel}'s view of the ``dusty'' side of the Universe.
\medskip

The survey will allow us to understand the most fundamental open
questions in galaxy formation today: the connection between the star
formation and the mass assembly history of galaxies; the interplay of
star formation and activity from a central supermassive black hole in
shaping the properties of galaxies; the connection between chemical
composition and the formation histories of galaxies (``extragalactic
arch\ae ology''); and the contribution of the environment and the
pervading dark matter halos to the formation of galaxies.

The main science questions that the mission will answer are: 
\begin{compactitem}
\item The connection between the star formation history and the mass assembly history.
\item The role of AGN and supernova feedback in shaping the 
  formation histories of galaxies, with a quantitative estimate of quenching timescales.
\item The formation of the first galaxies.
\item The source of reionization.
\item Evolution of the metallicity-mass relation, including [$\alpha$/Fe] and
  individual abundances.
\item Initial Mass Function as a tracer of star formation modes.
\end{compactitem}
\vskip+1mm
\phantom{.}
\end{NR}

\clearpage

\begin{NR1}[userdefinedwidth=17.5cm]
\section{A natural choice for the next L-class mission}
\label{sec:Intro}
\end{NR1}

\begin{multicols}{2} 

\begin{figure*}
\begin{center}
\raisebox{-0.5\height}{\includegraphics[width=55mm]{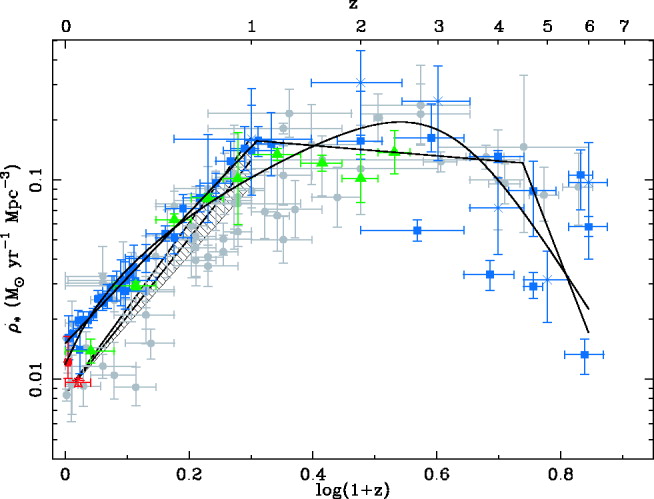}}
\hspace{5mm}
\raisebox{-0.5\height}{\includegraphics[width=55mm]{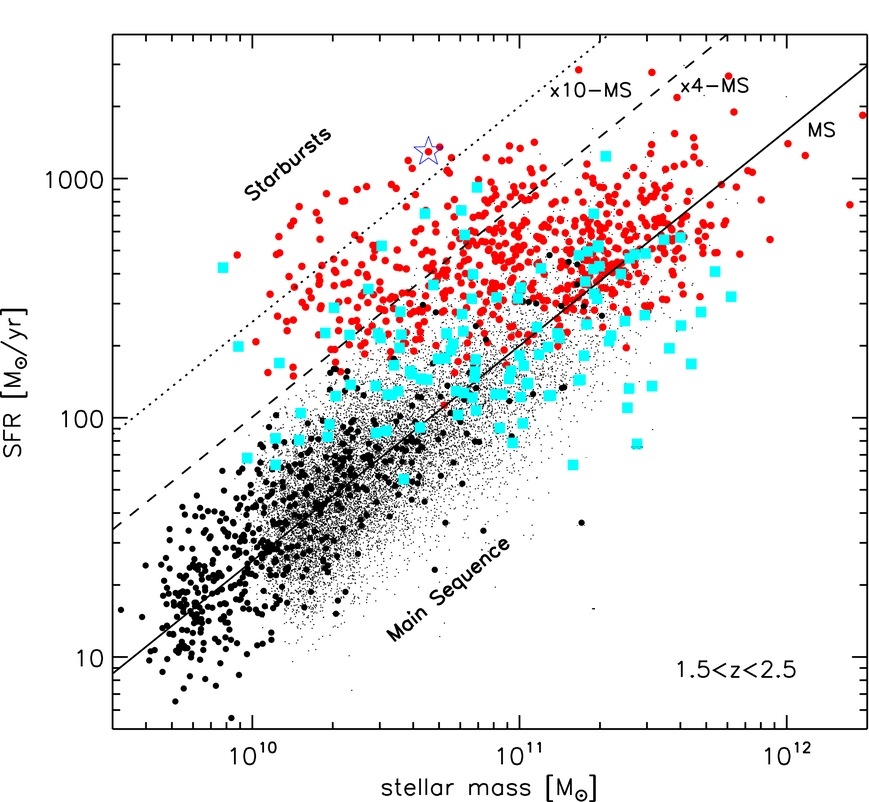}}
\hspace{5mm}
\raisebox{-0.5\height}{\includegraphics[height=75mm]{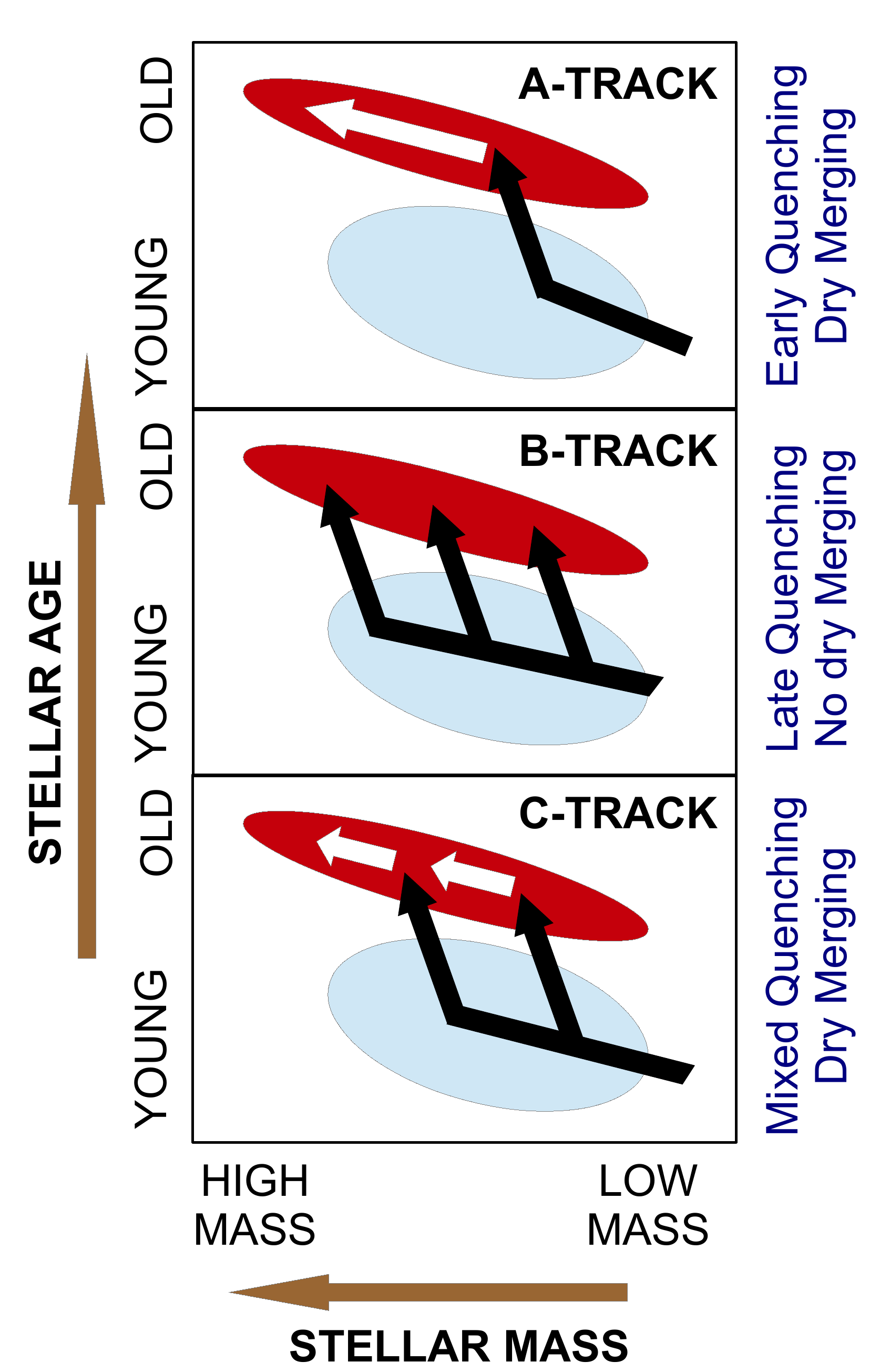}}
\end{center}
\caption{\small {\bf Left:} Cosmic star formation history \citep[from][]{HB:06}.
{\bf Middle:} Different modes of star formation \citep[from][]{Rodh:11}).
{\bf Right:} Schematics of galaxy evolution from the blue cloud to the red sequence
\citep[adapted from][]{Faber:07}.}
\label{fig:SFH}
\vspace{-0.3cm}
\end{figure*}

Even though astronomy as a human activity goes back to the origins of
civilization, and the understanding of celestial bodies has been
responsible for the scientific method that led to our technology-based
society, some of the most fundamental aspects of this scientific
discipline remain largely unsolved.  It was less than a century ago
that we began to understand the true meaning of galaxies as ``Universe
islands'', and even more recently that we could put them in context
with the underlying dark matter backbone and the evolution of the
Universe as a whole. The luminous component of galaxies is dominated
by stars, gas and dust. This L-class mission is designed to address
one of the major questions in ESA's ``{\sl COSMIC VISION 2015-2025}''
BR-247 document, namely question 4: {\bf How did the Universe
  originate and what is it made of?}, and especially topic 4.2: {\bf
  The Universe taking shape}.
\medskip

Two of ESA's recent cornerstone missions concentrate on very specific --
and essential -- aspects of this problem, namely the evolution of the
dust in star forming galaxies ({\sl Herschel}) and the history of our
own Milky Way galaxy ({\sl Gaia}).  The global structure and evolution
of the Universe is the main target of ESA's {\sl Planck} and {\sl
  Euclid} missions. Now, the next Large mission
should tackle the much wider problem of understanding the ``baryon
physics'' of galaxy formation, namely the highly complex set of
physical mechanisms responsible for the transformation of the
primordial hydrogen and helium gas mixture into the galaxies we see
today.
At a more fundamental level, it is possible to understand the
nature of dark matter only if we have a comprehensive understanding
of galaxy formation.
Such an endeavour requires the highest quality spectroscopic dataset
over a large survey, probing two major cosmic epochs:
a) the epoch at the peak of galaxy formation activity, namely between
redshifts z$\sim$3 and 1 (i.e. 2 and 6 billion years after the Big
Bang), and
b) the formation of the first galaxies, at redshifts between z=12 and
6 (between 400 and 900 million years after the Big Bang).
The spectroscopic analysis of galaxies constitutes the equivalent of
DNA fingerprinting, allowing us to determine the composition of the
galaxy and its past formation history. 

Even though ongoing and future surveys, such as {\sl Euclid}, {\sl
  WFIRST}, {\sl LSST} or {\sl SKA}\footnote{Although {\sl SKA} will
deliver an unprecedented map of the first stages of structure formation,
its design will only target the gas component through observations of the
H\textsc{I} 21\,cm line.} will give insightful clues about
the origins of galaxies, high quality spectroscopic data at moderately
high resolution ($R\equiv\lambda/\Delta\lambda\approx 1,000-2000$) is
the {\sl only} way that this problem can be solved.  While
ground-based NIR spectroscopy will undoubtedly make important strides
in the coming decades, OH and other backgrounds inherent to ground
based NIR spectroscopy will ensure that the spectroscopic performance
of \chronos -- at the faintness levels and spectral coverage required
for this science -- will remain unchallenged.  The behemoths of
astronomical science in the coming decades ({\sl JWST} from space or
the {\sl E-ELT} from the ground) will, in all likelihood, target in
detail specific issues of this project, but their very limited field
of view makes them incapable of gathering anywhere near the scale of
dataset required to make the extensive advances \chronos will deliver
in the area of galaxy formation and evolution. Rather than a
competitor, ultra-large observing facilities will be a complement to
\chronos.

\myfigure{
\begin{NR2}[userdefinedwidth=85mm]
\begin{center}
\vspace{-5mm}
\underline{\chronos in a nutshell}
\end{center}
\chronos is a dedicated 2.5m space telescope optimized for
ultra-deep NIR spectroscopy at moderate resolution (R=1500) in the
0.9-1.8$\mu$m range. The 5-year long, two-tiered survey will reach
H$_{\rm AB}$=26 over a 100\,deg$^2$, and H$_{\rm AB}$=27.2 over 10\,deg$^2$ at a
$5\sigma$ level in the continuum. The two main science drivers are the
formation of galaxies at the peak of activity (1$<$z$<$3) and
the first galaxies and the source of reionization (z$>$6).
\end{NR2}
}

\end{multicols}

\begin{NR1}[userdefinedwidth=17.5cm]
\section{The evolution of galaxies at the peak of activity}
\label{sec:D4000}
\end{NR1}

\begin{multicols}{2} 

\subsection{Introduction}

Over the past two decades, advances in detector technology have
allowed us to probe the evolution of star formation with cosmic time
(Fig.~\ref{fig:SFH}, left panel).  Between the formation of the first
galaxies and the present time, there was an epoch when the global star
formation history was at its peak, in the redshift interval between
z$\sim 3$ and $1$.  During this epoch, the majority of the stars in
the present Universe were formed and assembled into galaxies, making
it -- along with the very first epoch of galaxy formation (see
Sec.~\ref{sec:Hiz}) -- the most important interval of cosmic
history. In the ``local'' Universe (i.e. z$\simlt$0.1), large
spectroscopic surveys, most notably the Sloan Digital Sky Survey
\citep[hereafter SDSS,][]{SDSS} provided enough data to trigger a
quantum leap in our understanding of galaxy formation. In the
post-SDSS era it is possible to dissect datasets according to various
properties such as stellar mass, velocity dispersion, morphology,
environment, enabling us to ``ask the right questions''. SDSS has
shown that gathering a {\sl complete} database comprising up to a
million spectra is necessary to constrain in detail the many aspects
that contribute to the process of galaxy formation.  The spectroscopic
analysis of the stellar populations in the SDSS Universe 
confirmed in exquisite detail the strong bimodality between passive, and
predominantly massive galaxies (the so-called {\sl red sequence}) and
a {\sl blue cloud } of star forming systems \citep{Baldry:04}.
Furthermore, SDSS also revealed the presence of a characteristic mass
scale in the local Universe, around $3\times 10^{10}$ solar masses in
stars \citep{Kauff:03}, which marks a clear transition in the baryon
content of galaxies \citep{Moster:10}, reflecting different modes of
formation, either at a fundamental level down to star forming regions,
and/or at a global level reflecting the contribution from a
supermassive black hole, from feedback associated to star formation,
or even from the environment where galaxies live.

Star formation in galaxies seems to have two possible channels:
``normal'' star forming galaxies, such as our own Milky Way galaxy;
and so-called starburst galaxies, where an intense rate of
formation implies a much higher efficiency in the conversion of gas
into stars.  Fig.~\ref{fig:SFH} (middle panel) shows that galaxies in
the 1$<$z$<$3 redshift window have a wide range of star formation
efficiencies, from the ``quiescent'' main sequence phase to intense
starbursts. Various processes involving star formation, quenching and
mergers have been invoked to explain the observed trends
(Fig.~\ref{fig:SFH}, right panel). A very large number of papers have
been devoted to this open problem, at the observational, theoretical
and modelling levels. However, a definitive answer beyond simple
sketches of the evolution is possible only with detailed spectroscopic
information about the stellar content of galaxies over the peak of
formation activity, i.e. z$\simlt$3.  \chronos will obtain
detailed formation histories over the range of stellar mass,
redshift, and environment, required to decipher the mechanisms that
control the efficiency of star formation.

\myfigure{
\begin{NR2}[userdefinedwidth=85mm]
\begin{center}
\vspace{-5mm}
... in a nutshell
\end{center}
$\bullet$ Complete sample of galaxies out to z$\sim$3
down to $10^{10}$M$_\odot$ in stellar mass.\\
$\bullet$ Accurate assessment of environment over
the most active period of galaxy formation, probing
the interplay between dark matter and baryons.\\
$\bullet$ Detailed age, metallicity, abundance ratios, IMF: 
extragalactic arch\ae ology.\\
$\bullet$ Understanding the mechanisms controlling the growth of
galaxies: infall, outflows, AGN feedback, supernov\ae-driven winds.
\end{NR2}
}

\begin{figure*}
\begin{center}
\includegraphics[width=140mm]{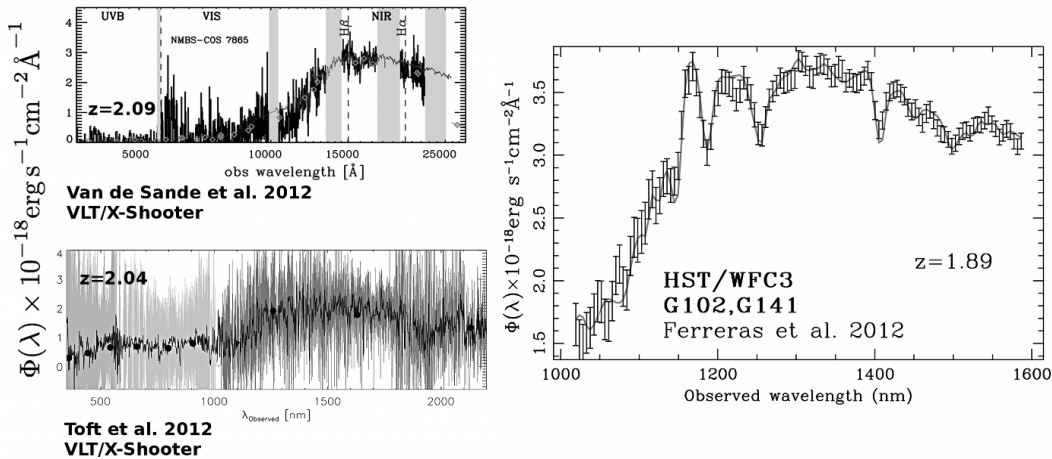}
\end{center}
\caption{\small Comparison of NIR spectra from the ground (VLT/X-Shooter, left,
  comprising integration times of 5-7 hours) and from space
  (HST/WFC3 slitless grisms, with an integration of just over 1 hour).
  These galaxies are very massive, with an apparent magnitude H$_{\rm AB}\sim$20.
  \chronos\ will extract $\sim$1--2 million spectra down to H$_{\rm AB}\sim$25-26.}
\label{fig:NIRSed}
\vspace{-0.3cm}
\end{figure*}

\subsubsection*{\vskip-8mm Why do we need a new spectroscopic survey?}

In the local Universe, SDSS provides detailed high quality
spectroscopic information only out to a relatively modest apparent
magnitude, and samples used for detailed spectroscopic analysis are
often restricted to z$\simlt$0.1. In addition, the passband-shifting
effect as we move into high redshift implies that the region around
the 4000\AA\ break -- which is highly sensitive to the properties of
the stellar populations -- moves into the Near Infrared (NIR), so that
future spectroscopic surveys in the optical region, such as MS-DESI
(restricted to $\lambda<1$\,$\mu$m) will not be able to target galaxy
formation at the peak of activity. In the NIR, ground-based
observations are hampered by the high atmospheric
background. Fig.~\ref{fig:NIRSed} compares state of the art
ground-based NIR spectroscopy of z$\sim$2 galaxies taken by the
X-Shooter instrument at ESO's {\sl Very Large Telescope} with a simple
slitless grism spectrum taken by the WFC3 on board the {\sl Hubble
  Space Telescope} at a similar redshift. The ground-based data was
obtained at a higher spectral resolution, nevertheless, the difference
in SNR is remarkable. Even though the field of view and spectral
resolution of the {\sl HST} data falls far below our target
specifications, the figure illustrates that a 2.5m telescope in space
is capable of superb deep NIR spectroscopy unrivalled from the
ground. In addition, the restriction of ground-based observations to
the allowed bands $J$,$H$,$K$ introduce ``redshift gaps'' that will
prevent a comprehensive study. Quoting \cite{SM:12}: ``Ultimately, one
needs a spectroscopic survey akin to SDSS at z=1$-$2''. However, {\sl
  even with the large field of view provided by Subaru's prime focus
  instruments (e.g. PFS), the quality required for a comprehensive analysis of
  the stellar populations of galaxies over the z$\sim$1$-$3 range
  requires a space telescope.}  Over a 5 year period, \chronos
will deliver millions of high quality spectra plunging down to a flux
level between 100 and 500 times lower than those shown in
Fig.~\ref{fig:NIRSed}. In the future, observatories such as the {\sl
  E-ELT} from the ground or {\sl JWST} from space will be capable of
achieving such low flux levels. However, the very small field of view
covered by these facilities will make surveys over many square degrees
unfeasible. 
\emph{As a consequence, neither {\sl E-ELT} nor {\sl JWST} will be
  capable of investigating the large scale environment of high
  redshift galaxies}.
\chronos will be the only facility able to provide a large dataset of
deep, high quality spectroscopic data in the NIR over large areas of
the sky, required to properly assess the role of environment on the
physical properties of galaxies and on their evolution.
Neither broad-band nor medium-band photometric surveys such as DES,
J-PAS or LSST can give enough ``spectral resolution'' to answer the
key open questions of galaxy formation and evolution. Even at moderate
resolution (e.g. ESA's {\sl Euclid} -- and possibly NASA's {\sl
  WFIRST} -- will provide $R\simlt 600$ slitless grism spectroscopy,
where the effective resolution is limited by the extent of the surface
brightness profile of the galaxy), it will not be possible to obtain
accurate constraints on the processes underlying the formation of
stars in galaxies. Furthermore, as cosmology-orientated surveys, they
are not designed to achieve high enough signal-to-noise ratio in the
continuum for faint sources, a strict requirement in our
mission. Nevertheless, for our purposes, {\sl Euclid} is, rather, a
valuable complement, as it will greatly help in the selection of
targets for detailed spectroscopy with \chronos, providing in
addition morphological information and photometry in several bands.

\subsection{Probing galaxy formation through their stellar content}

The redshift range of z$\sim$1$-$3 is a fundamental
epoch of galaxy formation for several reasons:
\begin{compactitem}
\item[(A)] It is the peak of the cosmic star formation history \citep{HB:06}.
\item[(B)] It is the peak of the AGN activity \citep{2006AJ....131.2766R}.
\item[(C)] It is the peak in the merger rate  \citep{2008ApJ...678..751R}. 
\item[(D)] It is the epoch when hosting haloes of massive galaxies
  allows for cold accretion via cosmic streams
  \citep{2009Natur.457..451D}. 
\end{compactitem}

Looking at the star formation history of galaxies today via moderately
high resolution spectra reveals the integrated star formation history
of these galaxies over all their progenitors
\citep{2005ApJ...621..673T,dlR:11}. The hierarchical paradigm of
structure formation predicts that the number of such progenitors can
be quite significant \citep[e.g.,][]{2006MNRAS.370..902K}. However, based on
observational constraints at z=0, it is not possible to estimate the
importance of the role of such progenitors and hence the role of
merging. This is mainly due to the 'cosmic conspiracy' of the star
formation main sequence, which, to first order, shows a linear
relation between star formation rate and stellar mass of the galaxy
\citep{2007ApJ...670..156D}. Thus, by knowing the star formation rate
of a present-day galaxy at any higher redshift, it is not possible to
determine whether it formed all its stars in one main progenitor or in
many. The {\sl only} viable option to observationally relate the star
formation history and the mass assembly history involves deep
spectroscopic observations, probing the underlying stellar populations
during the peak of activity. A complete mass-limited sample will serve
as an fundamental, unbiased benchmark to relate galaxies at different
redshifts with their merging histories. 

\chronos will provide a definitive statement on the formation
timescale of galaxies with respect to morphology, mass and
environment. It will also deliver information about the velocity
dispersion and chemical enrichment of the populations. Such studies
not only constrain but also motivate significant developments in
numerical simulations in a cosmological context, to achieve a more
consistent view of how galaxies form and evolve. For instance, the
evidence that massive galaxies are old and enhanced in Mg over Fe
\citep[e.g.,][and references therein]{Renzini:06} points towards an early
and rapid formation, thus constraining the timescales in which haloes
in regions of the Universe that are destined to form a cluster
collapse \citep[e.g.,][]{DeLuc:06}. A powerful test of these models is
to study differences in the stellar content of galaxies in different
environments. However, the evolutionary trends of stellar populations
can be hidden due to the age-metallicity degeneracy, which not only
affects the colours but also to a great extent the absorption
line-strength indices of the old stellar populations
\citep[e.g.,][]{Worthey:94}, if there is a relation between the age
and the metallicity of the galaxies \citep[e.g.,][]{FCS:99}. Much
progress has been made during the last decade to lift this degeneracy,
however, the chief aspect of this spectroscopic survey is that the
targeted redshift interval represents a range in lookback time that
resolves the ``age axis'' directly. Furthermore, studying the galaxies
when they were younger allows to derive much more accurate ages as, in
this regime, the spectral indicators have much larger variations for
smaller changes in the mean age \citep{Jorg:13}.

\begin{figure*}
\begin{minipage}[h]{\textwidth}
\centering
\raisebox{-0.5\height}{\includegraphics[width=65mm]{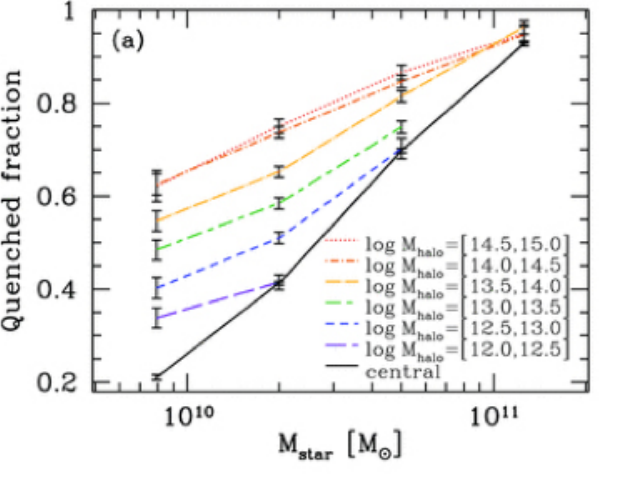}}
\raisebox{-0.5\height}{\includegraphics[width=95mm]{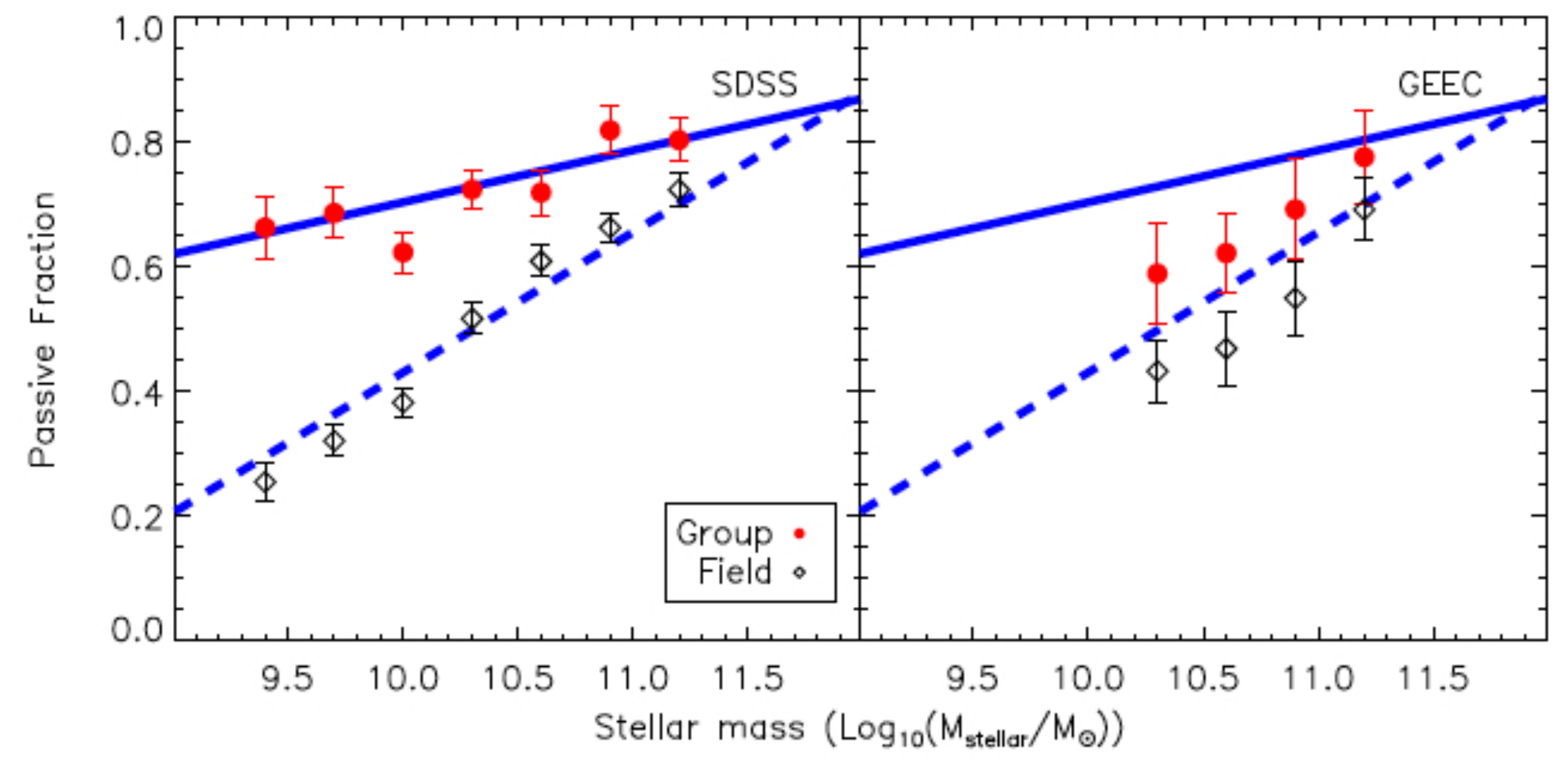}}
\end{minipage}
\caption{\small {\bf Left:} The fraction of quenched galaxies (whose specific star
  formation rate is lower than 10$^{-11}$ yr$^{-1}$) as a function of
  their stellar mass M$_{\star}$, for galaxies residing in
  environments of different content of dark matter (M$_{\rm halo}$). The
  solid line shows the variation of the fraction of quenched central
  galaxies (the most massive galaxy within each environment) in
  M$_{\star}$ \citep[from][]{Wetz:12}. {\bf Right:} The fraction of
  quenched galaxies as a function of their stellar mass in galaxy
  groups and in the field, as measured at z$\simeq$0 in the SDSS
  survey and at z$\simeq$0.4 in the GEEC survey
  \cite[from][]{McGee:11}.}
\label{fig:env1}
\vspace{-0.3cm}
\end{figure*}

Constraining the characteristic timescales for the formation of the
bulk of the stellar populations has been a major endeavour. This is
performed through the study of the chemical composition of galaxies
derived from their spectra. As different elements are released to the
interestellar medium by stars of different masses and, therefore, over
different timescales, stellar and gas abundance ratios (once suitably
calibrated) provide potential cosmic 'clocks' capable of eliciting the
timescale of star formation within a galaxy. It is important to
understand how non-solar abundance-patterns might affect the main
conclusions derived from the stellar population analysis. These
timescales can be fine tuned if, apart from [Mg/Fe], other abundance
ratios, such as [CN/Fe], are included in this analysis
\citep[e.g.,][]{Carr:04}. However, the different behaviour of elements released
by massive supernovae \citep[e.g.,][]{Woosley:02} remain unclear,
among other reasons, because these studies are still in their
infancy. The dearth of quality spectroscopic observational datasets
over a wide range of cosmic time explains the deficiencies of our
understanding on the distribution of chemical elements in galaxies.

\subsubsection*{\vskip-8mm Efficiency of star formation}

The efficiency of converting baryons into stars within given dark
matter haloes is of prime interest. Theoretical predictions of
LCDM-based models suggest a state of self-regulation in which the star
formation rate is controlled by the growth rate of dark matter haloes
\citep[see, e.g.,][]{Bower:06,2010ApJ...718.1001B,Bower:12,Guo:13}.
The unprecedented sample of galaxies that \chronos will provide
over the z$\sim$1--3 range will allow us to construct high precision
correlation functions for the galaxy population, including detailed
information about their star formation histories, relating galaxy
growth with the underlying distribution of dark matter structure. In
this way, it will be possible to link measured star formation
histories to theoretically predicted growth rates of dark matter
haloes. It has been shown in studies at low redshift, that by only
using abundance matching techniques it is not possible to obtain
robust constraints on the galaxy halo occupation function
\citep{2011MNRAS.416.1486N}.

\subsubsection*{\vskip-8mm Galaxy Formation and the Initial Mass Function}

One of the most fundamental properties of star formation, the stellar
Initial Mass Function (IMF), which describes the mass distribution of
stars at birth, is assumed to be universal and constant with cosmic
time. Recently, it has been argued that the stellar initial mass
function may not be universal; differences have been hinted in the
most massive galaxies at z$\simeq$0 \citep{vDC:10,Cap:12}.  A
combination of a large, high-quality spectroscopic dataset from SDSS
and detailed population synthesis models \citep{Vazdekis:12} enabled
the confirmation of a systematic trend with velocity dispersion
\citep{Ferreras:13,FLB:13}. Such non-universality reflects fundamental
differences in the mode of star formation with respect to galaxy mass
\citep{Hopkins:12}, that need to be addressed in ab initio simulations
of star formation \citep[e.g.][]{Bate:03}.  Furthermore, recent
theoretical developments suggest that the IMF shape and mass-cutoffs
might depend on the star formation rate \citep{Weidner:11}. However,
the imprints of a varying IMF on spectra might be coupled to a
variation of the abundance ratio of certain chemical species
\citep{CvD:12}. Future population synthesis models (see
\ref{subsec:pop}) along with high quality spectroscopic data of
galaxies probing a wide range of cosmic time will allow us to
disentangle these effects.  With \chronos, it will be possible to
probe the evolution of the IMF during the most important epoch of star
formation, spanning a range of mass, velocity dispersion and
metallicity. This issue is fundamental for an accurate assessment of
the cosmic star formation history -- which depends on the assumptions
made for the underlying stellar populations. Moreover, derived star
formation histories may have to be revised depending on these results,
as a systematic change in the IMF can affect the model predictions
relating the distribution of stellar ages and metallicities. We
emphasize that such studies require high quality NIR spectroscopic
data of very faint sources, such as those that \chronos will
provide, beyond the capabilities of any spectroscopic survey in the
coming decades.

\subsection{The ecology of galaxies}

In addition to the intrinsic mechanisms described above, the
environment where galaxies live plays a fundamental role in shaping
their evolution, as it is capable of quenching their star formation by
removing their hot and cold gas reservoirs and to literally disrupt
them by removing their stars.  From the observed properties of
galaxies at z$\simeq$0 we have collected a large body of evidence for
the occurrence of such environmental processes, but the determination
of their timescales and amplitudes remains at a qualitative level.  We
have not yet established in a quantitative way how these parameters
depend on environment and redshift, i.e. on the assembly history of
a galaxy cluster or galaxy group, through cosmic time.

The unprecedented statistical power of SDSS, in terms of the
photometric and spectroscopic properties of galaxies measured at
optical wavelengths, has allowed us to describe the behaviour of the
star formation activity of galaxies across many orders of magnitude
with respect to their stellar mass and environment at z$\simeq$0. We
know that the population of quenched galaxies -- not forming new stars
any longer -- increases with their stellar mass for a given kind of
environment, and with environment size (from small galaxy groups to
large clusters) at fixed stellar mass 
\citep[see Fig.~\ref{fig:env1}, left;][]{Wein:06,VdB:08,AP:09,Wetz:12}.

Galaxies become increasingly older (in terms of the mean age of their
stars) as their environment becomes more massive (from galaxy groups
to clusters), suggesting that galaxies in today's clusters were
accreted at earlier times (i.e. at a higher redshift of infall) than
galaxies in today's groups and had their star formation activity
suppressed for longer times \citep{AP:10}. Most likely the quenching of
their star formation activity happened while these galaxies were still
living in smaller groups, which merged at later times with bigger
structures like galaxy clusters.

Unfortunately, observations of z$\simeq$0 galaxies can not constrain
their redshifts of infall, or the time when they were subjected to
environmental effects for the first time. Both {\sl Euclid} and 
\chronos will trace the assembly history of environments with cosmic
time and provide us with a direct measurement of the redshift of
infall of galaxies as a function of their stellar mass.  In addition,
the lensing information from {\sl Euclid} will be combined with the
spectroscopic information produced by \chronos to probe the
dependence of the star formation histories on the dark matter halos.
However, while {\sl Euclid} will only trace the assembly of the very
massive end, with a significant bias towards star-forming galaxies,
\chronos will extend the study to smaller masses, including old
populations, and therefore, avoiding the selection bias of the {\sl
  Euclid} sample. A comparison of the z$\simeq$0 data with the
predictions of semi-analytic models of galaxy evolution indicates that
galaxies should quench their star formation over a few billion years;
this is the only available and indirect estimate of the timescale for
environmental quenching of star formation and it is unclear how much
it depends on environment and whether it has changed with
redshift. When and in which environments did the quenching of the star
formation activity of galaxies start?  How fast did it proceed?  The
quantitative and direct answers to these questions come from the
measurements of star formation rates, star formation histories and
chemical enrichment of galaxies of different stellar mass, in
different environments at different epochs, from z$\sim$1--3 to
z=0. Only these observables provide a direct estimate of the typical
timescales of star formation in galaxies and hence a model-independent
estimate of the timescales with which different environments succeeded
in quenching their member galaxies of different stellar mass, and gave
rise to the present-day galaxy populations.

With increasing redshift such measurements move to infrared
wavelengths and become challenging even for modern ground-based
telescopes. The intervening Earth atmosphere offers us only a partial
disclosure of galaxies properties at z$>$0.5; we can mostly measure
emission lines (hence star formation rates), while absorption lines
(age and metallicity indicators) become less and less accessible.
From the data collected so far on galaxies at 0.3$<$z$<$0.8, we know
that the fraction of quenched galaxies is larger in galaxy groups than
in the field, but defintively lower than the fraction of quenched
galaxies in groups at z$\simeq$0 
\citep[see Fig.~\ref{fig:env1}, right;][]{Wil:05,McGee:11}. At intermediate
redshifts, the fraction of star forming galaxies decreases from 70-100\%
in the field to 20-10\% in the more massive galaxy
clusters \citep{Pogg:06}. Nevertheless, in terms of their star
formation rates, galaxies in groups are not significantly different
from those in the field; only star forming galaxies in clusters
exhibit star formation rates a factor of 2 lower than in the field at
fixed stellar mass \citep{Pogg:06,Vul:10,McGee:11}.

In the highest redshift range probed for environment at present,
0.8$<$z$<$1, the more massive galaxy groups and clusters are populated
mostly by quenched galaxies along with a 30\% fraction in
post-starburst galaxies \cite[i.e. with a recently truncated star formation
  activity;][]{Bal:11}.  The fraction of post-starburst galaxies is a
factor of 3 times higher in clusters than in the field. Cluster and
field galaxies are instead very similar in terms of the strength of
their star formation activity and the amount by which their star
formation has been quenched. These results have led \citet{Muzz:12} to
conjecture that either the quenching of star formation due to the
secular evolution of galaxies dominates over the quenching induced by
galaxy environment, or both mechanisms occur together with the same
timescale.  Which timescale?  We do not currently know.  In order to
make further progress, we require a facility such as \chronos to
observe a complete stellar-mass limited sample of environments at
z$\simgt$1, and to measure the star formation histories of their
galaxies with an unprecedented accuracy, thus providing the fading
timescales of star formation of galaxies of different stellar mass
inhabiting different environments. This is not simply an incremental
step in our knowledge of environment-driven galaxy evolution. This is
the {\sl fundamental quantitative change} from the simple head-count of
quenched or star forming galaxies to the measurement of physical
properties of galaxies in environments at z$\simgt$1, during the peak
of galaxy formation activity. Such a step makes it possible to compare
for the first time the same physical properties of galaxies at fixed
stellar mass in different environments between z=0 and z$\simgt$1, and
to firmly quantify the extent to which environment regulates and
modifies galaxy evolution across cosmic time.

\begin{figure*}
\begin{center}
\raisebox{-0.5\height}{\includegraphics[width=85mm]{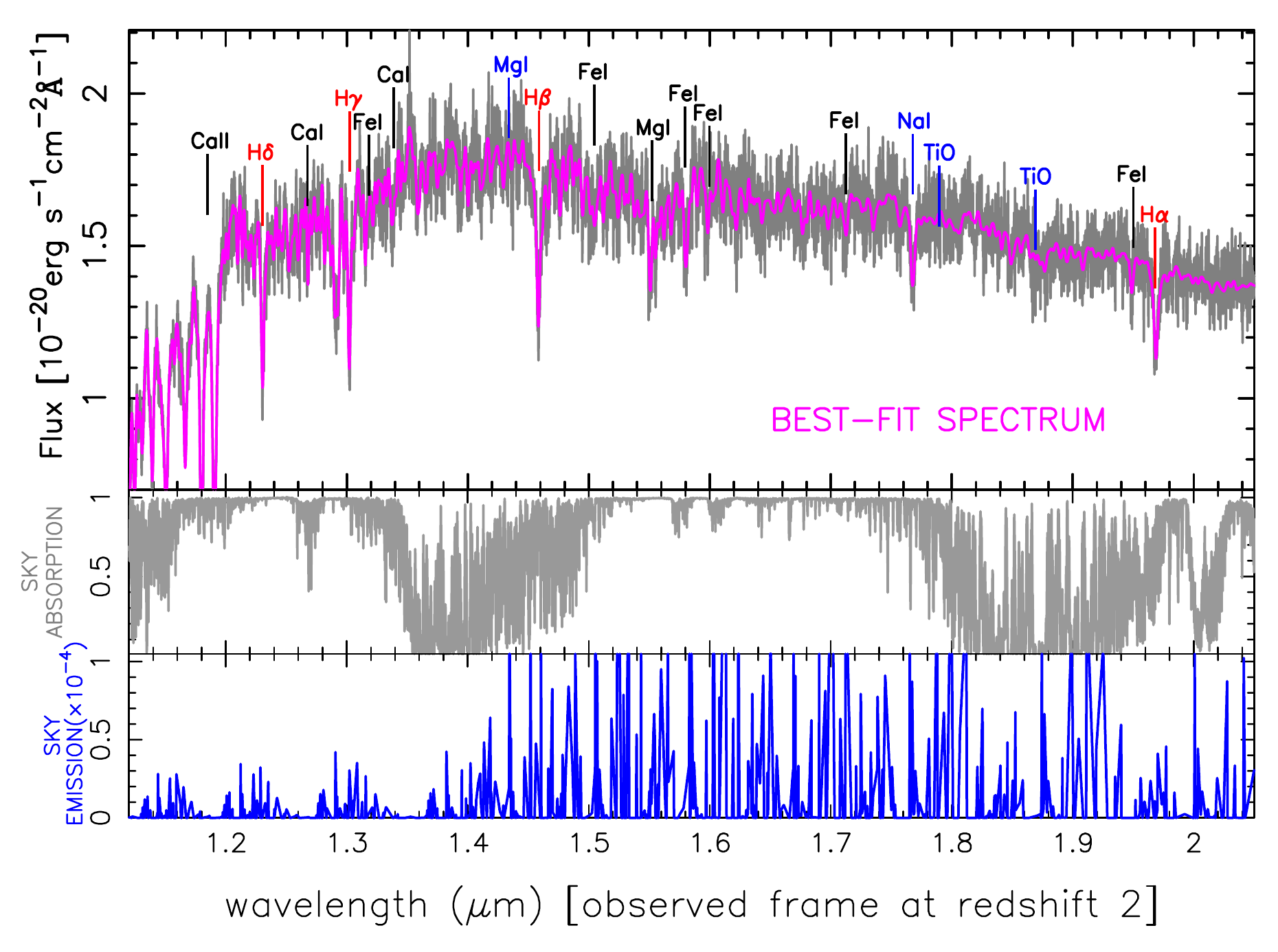}}
\raisebox{-0.5\height}{\includegraphics[width=75mm]{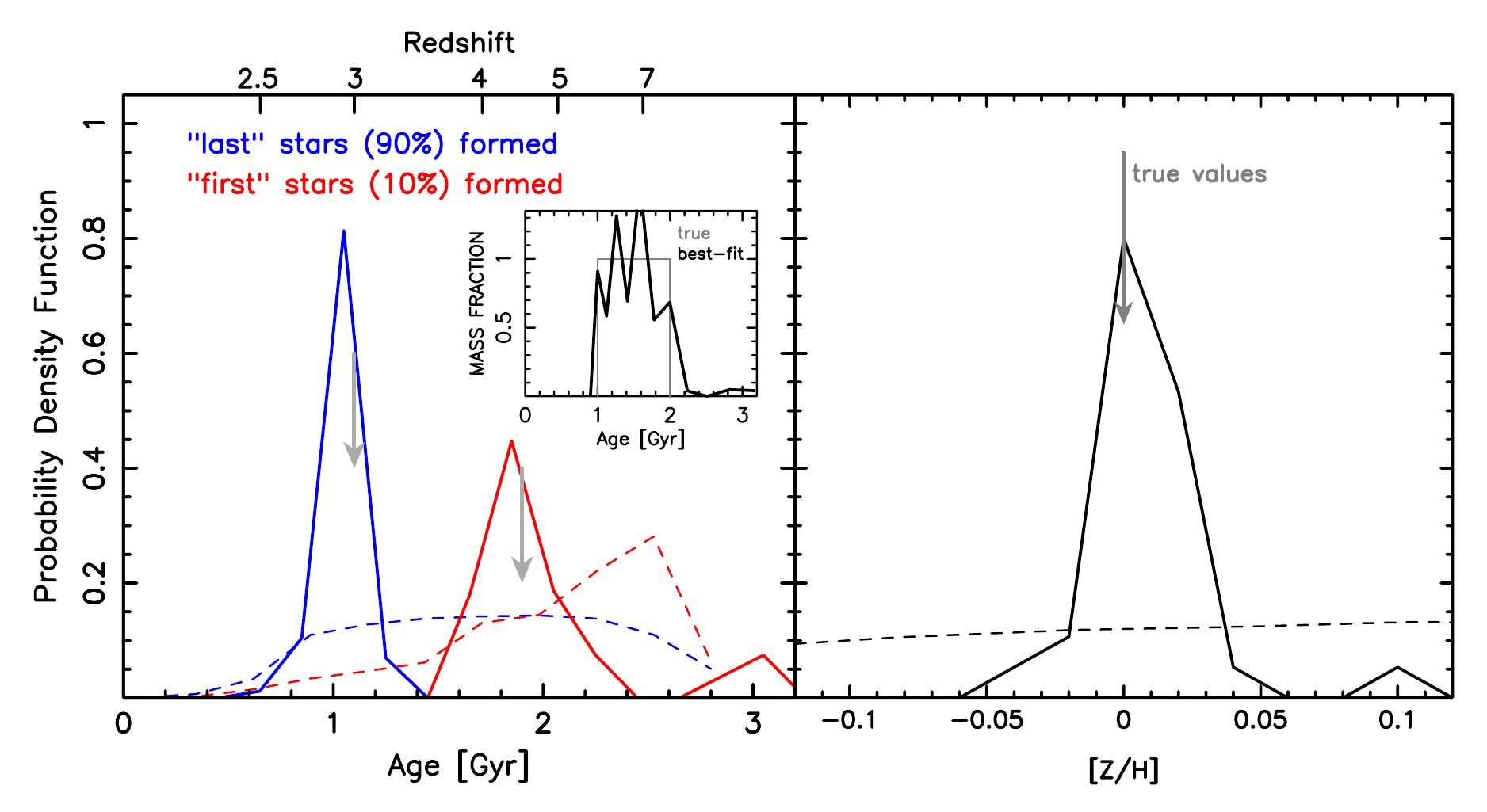}}
\caption{\small {\bf Left-Top:} Synthetic spectrum resembling the
  progenitor of a nearby early-type galaxy, with a stellar mass of
  $3 \times 10^{10} \, M_\odot$, at z$\sim$2 with a SNR of $20$ per
  resolution element (grey).  The best-fit model -- obtained by direct
  spectral fitting -- is overplotted in magenta.  Some spectral
  features, sensitive to the star formation history (red),
  metallicity/abundances ratios (black), and IMF (blue) are
  included. {\bf Left-Bottom:} sky emission (blue) and telluric
  absorption (grey), make the target spectrum very challenging from
  the ground.  {\bf Right:} Constraints on the timescales of star
  formation (left) and metallicity (right), derived from the spectrum
  on the left. The vertical grey arrows mark the input values. On the
  left, marginalized probability distribution functions (PDF) of the
  first (blue) and last (red) time of star formation (corresponding to
  10\% and 90\% of total stellar stars formed). The inset shows the
  ``true'' star formation history of the system (grey histogram), and
  one typical best-fitting estimate (black curve). Note that a deep
  {\rm photometric} survey cannot constrain the SFH or the metallicity in
  detail (dashed lines in both panels)}
\label{fig:SEDFit}
\end{center}
\vspace{-0.3cm}
\end{figure*}

\subsection{Revealing the stellar population content of z$\sim$1--3 galaxies}
\label{subsec:SpFit}

Understanding the nebular and stellar population properties of high
redshift galaxies is an essential step towards a self-consistent
picture of galaxy formation and evolution. The study of strong
emission lines in the spectra of galaxies at  z$\simgt$1 has
recently led to important results on the gas-phase properties, like
the fact that metallicity exhibits a sharp transition towards subsolar
values at z$\simgt$2.5 \citep[e.g.,][]{Moller:13}, and gas-rich disks
are more dispersion dominated than in the nearby Universe
\citep[e.g.,][]{Forster:11}.  Both results point to a major role of the
accretion of unprocessed gas during the assembly of galaxies. In stark
contrast, little is known about the stellar population content (i.e.
the overall star formation history, metallicity, and IMF) of galaxies
at z$>$1.  

\subsubsection*{\vskip-8mm Extracting star formation histories from spectra:\\ An example}

Fig.~\ref{fig:SEDFit} (left) illustrates the reason for this impasse,
that will render problematic our understanding of galaxy formation and
evolution in the coming two decades.  The upper panel plots a
synthetic model spectrum, resembling the progenitor of a nearby
early-type galaxy with mass $3 \times 10^{10}$M$_\odot$, as seen at
z$=$2 (grey spectrum).  The bottom panel shows the typical emission
spectrum from the night sky (blue) as well as telluric absorption
(grey). Such a high background -- several orders of magnitude higher than the
signal -- makes the spectrum intrinsically
inaccessible from a ground-based observatory, regardless of its photon
collecting power (see also Fig.~\ref{fig:NIRSed}).  The synthetic
spectrum of Fig.~\ref{fig:SEDFit} is obtained as a linear
combination of simple stellar population models from the MILES
synthetic library~\citep{Vazdekis:10}, assuming that the galaxy starts
forming stars at z$\approx$5 (corresponding to an age of $\sim 2$\,Gyr
at z$=2$), at a constant rate, with solar metallicity and a
Kroupa-like IMF, down to z$\sim 3$ (Age$\sim 1$\,Gyr), when star
formation is suddenly quenched (because of, e.g., internal and/or
environmental processes).  The SNR of the spectrum ($\sim 20$, per
resolution bin at R=1500) corresponds to a deep exposure, as planned
for the \chronos ultra-deep survey (see Sec.~\ref{sec:survey}).
Notice the strong Balmer lines in the spectrum (e.g.
H$\beta$ and H$\delta$), that reflect the recent ($\sim 1$\,Gyr)
quenching of star formation -- a fact only recently observed at
z$\sim$1.5~\citep{Bezanson:13,Ferreras:13}, and eventually
attributable to the presence of an AGN through suitable diagnostic
lines that, at z$\simgt$1, are hard to observe from the ground
(e.g.  H$\alpha$ and the companion [N\textsc{II}] line). A survey like
\chronos is therefore required for the redshift range
corresponding to the peak of galaxy formation activity.

Several absorption lines can be measured in the spectrum, most being
sensitive to total metallicity and to the chemical abundances of
individual elements (e.g. Mg, Si, Ti, Ca, Na), and some of them also
to the fraction of dwarf-to-giant stars in the stellar IMF (e.g.  the
TiO features, see blue hatched regions).  While the measurement of
(total) metallicity and, to a lesser degree, that of [$\alpha$/Fe]
abundance ratio are a common practice in the case of low-redshift
galaxies \citep[e.g.,][]{Gallazzi:05, Thomas:11}, abundance estimates
of single chemical species at low-z has become feasible only recently
\citep[e.g.,][]{Johansson:12, Conroy:13}, thanks to the rapid
development of stellar population models and spectral fitting
techniques.  During the next two decades we will develop superb
stellar population models and software tools, to constrain also the
star formation histories, abundance patterns, and the stellar IMF in
high redshift galaxies, provided that their spectra will become
accessible.  Estimating metallicity and abundance ratios for galaxies
at z$>$1 would therefore give us crucial insights into the chemical
enrichment history of galaxies (infall versus outflows of cold gas,
and preferential loss of metals from supernov\ae-driven winds),
nucleosynthesis yields, and the time-delay distributions of different
types of SNe (e.g. Type Ia relative to core-collapse, driving the
[$\alpha$/Fe] of a stellar population at different epochs), as
absorption features at high redshift would reflect the abundance of
different elements by the time and immediately after they are
synthetized in a galaxy.

During the next decade, direct fitting of stellar population models
to data will likely become the ``standard'' tool to optimally extract
the spectral information, compared to other well consolidated
approaches like the analysis of line strengths (e.g.  Lick-system
indices).  The magenta curve in the top panel of
Fig.~\ref{fig:SEDFit} (left) shows the result of fitting the
ultra-deep-survey-like synthetic spectrum with a linear combination of $50$
simple stellar populations, with different ages and metallicities.  Notice that
internal reddening -- certainly important at high redshift -- is set
to be zero when synthetizing the spectrum, while it is included as a
free parameter in the fitting procedure.  The superb quality of the
fit will be typical for data of the quality we envisage for \chronos, 
with flux calibration accuracy  better than a few
percent.  

Fig.~\ref{fig:SEDFit} (right) illustrates the possibility of
constraining timescales and metallicity for targets in the ultra-deep
survey. One hundred noise realizations of the above mock spectrum are
fitted as shown in Fig.~\ref{fig:SEDFit}, deriving from each
best-fitting mixture some relevant, illustrative, parameters, i.e.
the first and last epochs of star formation as well as the average
metallicity.  The first and last epochs are defined by the times when
the system formed 10\% and 90\% of its stellar mass,
respectively. Hence, NIR spectra with the given combination of SNR and
resolution allows us to constrain sufficiently the formation
timescales of a stellar population at redshifts corresponding to the
peak of formation activity, as well as its total metal content (at
less than 10\,\%).

Velocity dispersion (not shown in the plot) can also be constrained
with a $<$10\,\% accuracy.  Notice that the last epoch of star
formation is connected to the quenching mechanism (e.g.  AGN and/or
environment), while the first epoch is ultimately driven by the
initial conditions of density pertubations, along with other subtle
physics (like reionization preventing star formation in small halos)
illustrating the possibility to finally understand galaxy evolution in
a full cosmological framework. In comparison, deep photometric surveys
will not be able to compete on this front: the dashed lines in
Fig.~\ref{fig:SEDFit} (right) show the constraining power when only
using broadband photometry, simulating data for the same galaxy, at
H$_{\rm AB}$=25, from a survey 1\,mag deeper than the {\sl Euclid} wide
survey, and using a wide photometric coverage: $rizYJH+K$.

\subsection{Towards the first galaxies}
\label{subsec:zgt3}

\chronos will be designed primarily in order to obtain a complete
mass-limited sample of galaxies down to $\sim 10^{10}$M$_\odot$ 
over the $1\simgt$z$\simgt 3$ redshift
interval.  In addition, this redshift range allows us to observe the
rest-frame spectral window around the 4000\AA\ region, a highly
sensitive area to the age distribution of stars and their chemical
composition. Beyond this, though, the capabilities of the
instrumentation also opens up the $3\simgt$z$\simgt 6$ redshift
range. Although at those redshifts it will not be possible to observe
mass-limited samples, \chronos will obtain SFRs from the 
NUV
emission, which -- complemented with stellar mass estimates from
additional photometry from {\sl Euclid} and future ground-based NIR
photometric surveys will give a snapshot of the evolution of the
efficiency of star formation between the first phases of galaxy
formation at z$\simgt$6 (the topic of the next section), and the epoch at the peak
of activity (this section), acting as a bridge between these two
fundamental stages of cosmic evolution.  In addition, rest-frame NUV
spectral features such as Mg$_{\rm UV}$ \citep{Daddi:05} will help
characterize the properties of the stellar populations, although with
a significantly lower precision with respect to the z$\simlt$3 sample.

\subsection{Population Synthesis in 2030}
\label{subsec:pop}

The most common methodology for deriving relevant stellar population
parameters from the integrated light of galaxies consists in
confronting observational data to predictions from stellar population
synthesis models \citep{Tins:80}. However this approach is hampered by
various fundamental degeneracies such as that between the age and the
metallicity \citep{Worthey:94}. There are also other limitations that
entangle derivations of burst-age and burst-strength \citep{LR:96}
or effects from the IMF \citep{Vazdekis:03}. Such degeneracies are
commonly tackled with targeted spectral indices, direct spectral
fitting, or a combination of both. However, as the quality of these
models rely on the employed ingredients, great efforts are being put
to develop stellar models and spectral libraries. This goal is being
achieved in part by means of new stellar evolutionary calculations,
with updated input physics, which might eventually include Helium and
atomic diffusion, and higher mass/metallicity/age resolution
\citep{Pietr:04}. Stellar libraries at moderately high spectral
resolution with varying abundance ratios, either theoretical
\citep{Coelho:05} or empirical \citep{Milone:11} as well as
theoretical stellar evolutionary tracks with varying element mixtures
\citep{Pietr:06} are being developed. These libraries will lead to new
generations of stellar population synthesis models that are better
suited to estimate the observed abundance patterns, including the
measurement of {\sl individual} abundance ratios, opening the field of
extragalactic arch\ae ology. This information puts the
mass-metallicity relation ``under the microscope'', allowing us to
quantify in detail the various aspects of galactic chemical
enrichment, including the effect of infall and outflows, and its
connection with environment \citep{KM:08}; or the dispersion of metals
into the intergalactic medium \citep{Pontz:08}. In addition,
developments in stellar libraries in the NUV \citep[e.g.,][]{NGSL}
will optimise the methodology to extract information from the
z$\simgt$3 sample (see~\ref{subsec:zgt3}).

The advent of models predicting galaxy spectra at moderately high
resolution \citep[e.g.,][]{BC:03,Vazdekis:10} has opened the
possibility of establishing robust constraints on the star formation
histories, via a variety of full spectrum-fitting methods
\citep[e.g.,][]{Koleva:08}. There is a growing body of publications
based on this approach as it allows us not only to attempt to estimate
the star formation history (see~\ref{subsec:SpFit}) but also to
interpret better the results based on line-strength indices, which are
more biased towards recent bursts and therefore hide the contributions
weighted by mass. These estimates are particularly relevant for
assessing the various mechanisms proposed for the assembly of galaxies
and the role of environment.

\subsection{Synergy with {\sl Herschel}}
\label{subsec:herschel}

The synergies with {\sl Euclid} are obvious, and throughout this white
paper there are abundant references to the use of the {\sl Euclid}
surveys to aid in the target selection, analysis and interpretation of
the data. We devote this subsection to another of ESA's flagship
missions: {\sl Herschel} has provided, for the first time,
efficient imaging of large areas of the sky in the far-IR window, from
70 to 500\,$\mu$m. In particular, the Herschel Multi-tiered
Extra-galactic Survey \cite[HerMES,][]{Hermes} -- the largest
project on Herschel at 900 hrs -- mapped over 70\,deg$^2$, tracing
dust-enshrouded star-formation sources during the peak of galaxy mass
assembly.  \chronos will be essential to take full advantage of
the legacy of the {\sl Herschel} surveys providing unique
spectroscopic follow-up. Indeed, a detection in the FIR implies large
amounts of dust emission, i.e. reprocessed light from the UV stellar
emission. As a result the optical/near-IR SEDs are often very red and
a large fraction of luminous Herschel galaxies are very faint or
undetected in the optical bands, requiring deep NIR spectroscopy for
their redshift measurements.  However, over 90\% of the
250\,$\mu$m-detected sources have a counterpart at K$_{\rm AB}<$24. The
\chronos surveys are thus optimally designed to exploit in full
the investment of ESA in HerMES, providing redshifts, dynamical
masses, stellar population properties, local environment and
clustering for essentially all of the sources detected in the HerMES
survey.

\end{multicols}

\begin{NR1}[userdefinedwidth=17.5cm]
\section{Cosmic Reionization \& galaxy/black-hole formation}
\label{sec:Hiz}
\end{NR1}

\begin{multicols}{2} 

\subsection{Introduction}

Cosmic reionization is a landmark event in the history of the
Universe.  It marks the end of the ``Dark Ages'', when the first stars
and galaxies formed, and when the intergalactic gas was heated to tens
of thousands of degrees Kelvin from much lower temperatures.  This
global transition, during the first billion years of cosmic history,
had far-reaching effects on the formation of early cosmological
structures and left deep impressions on subsequent galaxy and star
formation, some of which persist to the present day.

The study of this epoch is thus  a key frontier in
completing our understanding of cosmic history, and is currently at
the forefront of astrophysical research \citep[e.g.][]{Robertson:13}.
Nevertheless, despite the considerable recent progress in both
observations and theory (e.g. see recent reviews by
\citealt{Dunlop:13} and \citealt{Loeb:13}) all that is really
established about this crucial era is that Hydrogen reionization was
completed by redshift $z \simeq 6$ (as evidenced by high-redshift
quasar spectra; \citealt{Fan:06}) and probably commenced around $z \sim
15$ (as suggested by the latest WMAP9 microwave polarisation
measurements, which favour a `mean' redshift of reionization of $10.3
\pm 1.1$; \citealt{Hinshaw:13}). However, within these bounds the
reionization history is essentially unknown, and new data are required
to construct a consistent picture of reionization and early galaxy
formation/growth.

Unsurprisingly, therefore, understanding reionization is one of the
key science goals for a number of current and near-future large
observational projects. In particular, it is a key science driver for
the new generation of major low-frequency radio projects (e.g. {\sl LOFAR},
{\sl MWA} and {\sl SKA}) which aim to map out the cosmic evolution of the 
{\it neutral atomic} Hydrogen via 21-cm emission and absorption. However,
such radio surveys cannot tell us about the sources of the ionizing
flux, and in any case radio observations at these high redshifts are
overwhelmingly difficult, due to the faintness of the emission and the
very strong foregrounds.  It is thus essential that radio surveys of
the neutral gas are complemented by near-infrared surveys which can
both map out the growth of ionized gas, and provide a complete census
of the ionizing sources.  A genuinely multi-wavelength approach is
required, and cross-correlations between different types of
observations will be necessary both to ascertain that the detected
signals are genuine signatures of reionization, and to obtain a more
complete understanding of the reionization process.

It has thus become increasingly clear that a {\it wide-area},
{\it sensitive}, {\it spectroscopic} near-infrared survey of the
z=6--12 Universe is required to obtain a proper understanding of
the reionization process and early galaxy and black-hole
formation. Such a survey cannot be undertaken from the ground, nor
with {\sl JWST} (inadequate field-of-view), or {\it Euclid}
(inadequate spectroscopic sensitivity). Only a mission such as 
\chronos can undertake such a survey and simultaneously address
the three, key, interlated science goals which we summarize below.

\begin{figure*}
\begin{center}
\includegraphics[width=50mm]{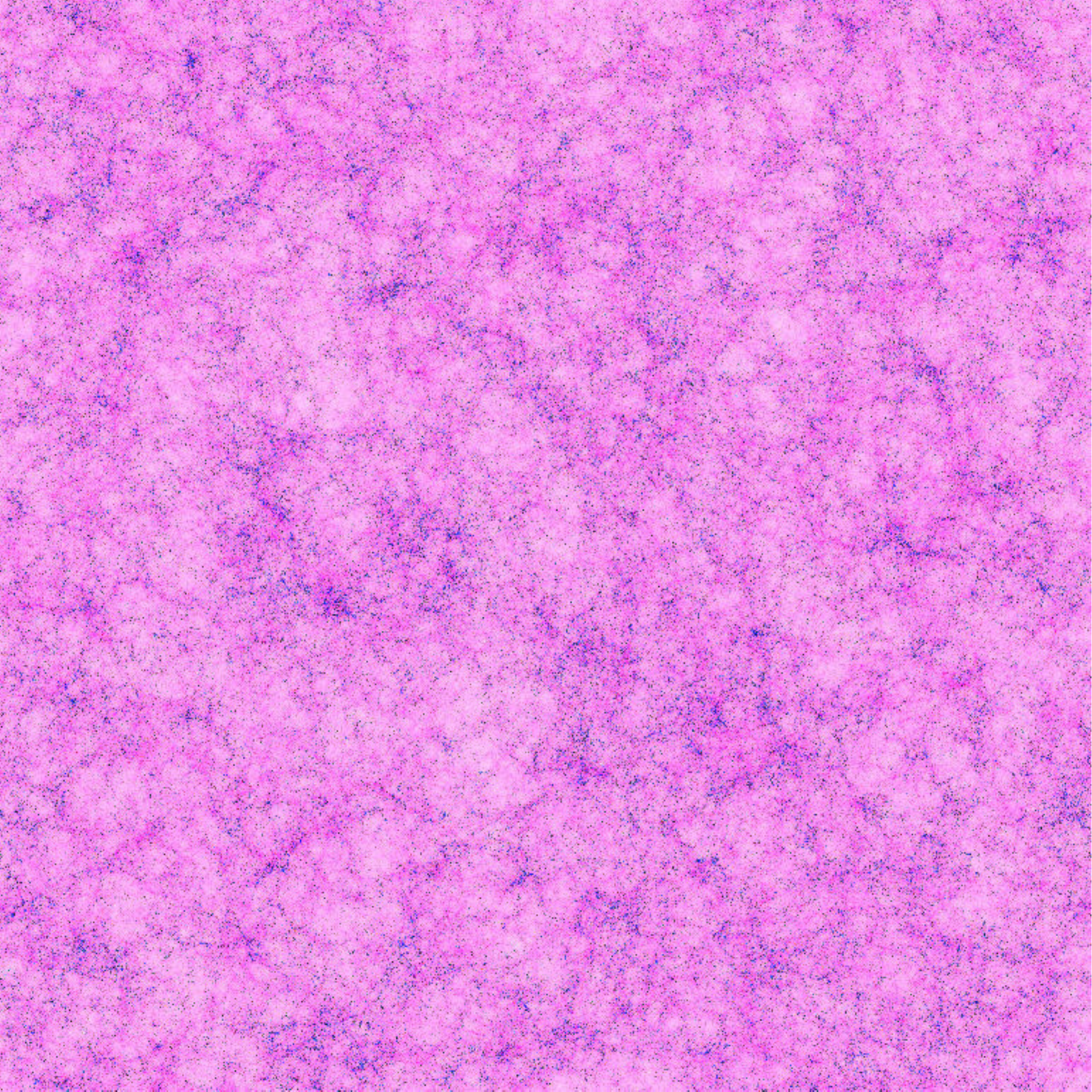}
   \hspace{8mm}
\includegraphics[width=50mm]{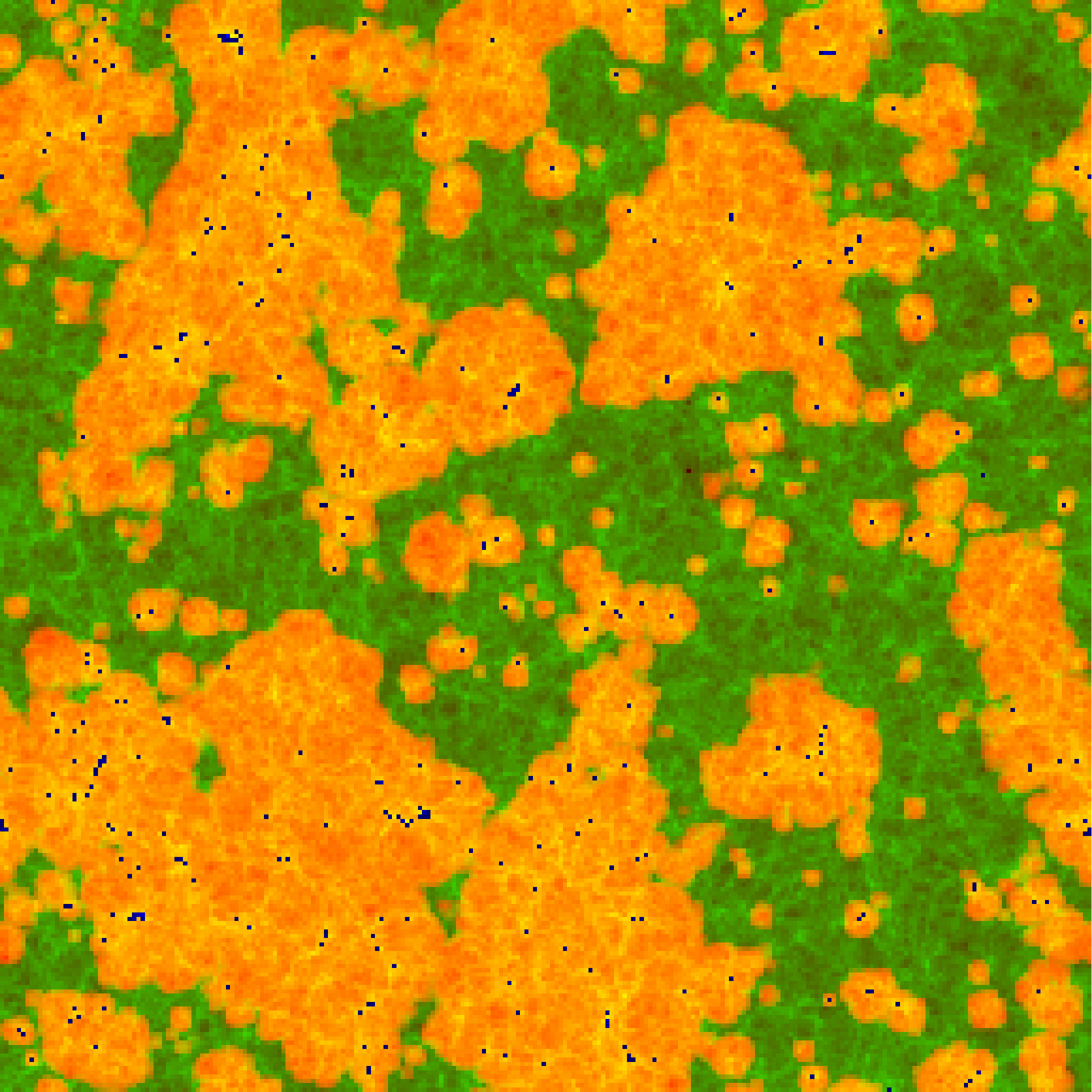}
\caption{\small {\bf Left} Early structure formation in $\Lambda$CDM
  (at $z=6$) from an N-body simulation with $5488^3$ (165 billion)
  particles and a volume $425\,\rm h^{-1} Mpc^3$. Shown are the
  dark-matter density (pink) and halos (blue).  This synthetic image
  corresponds to $3.5 \times 3.5$ degrees on the sky.  {\bf Right:}
  The geometry of the epoch of reionization, as illustrated by a slice
  through a $(165\,\rm Mpc)^3$ simulation volume at $z=9$. Shown are
  the density (green/yellow), ionized fraction (red/orange), and
  ionizing sources (dark dots) \citep{Iliev:12}.  The necessity of
  a deep, near-infrared spectroscopic survey covering many square
  degrees is clear.}
\label{fig:reioniz}
\end{center}
\vspace{-0.3cm}
\end{figure*}

\myfigure{\begin{NR2}[userdefinedwidth=80mm]
\begin{center}
\vspace{-5mm}
... in a nutshell
\end{center}
$\bullet$ Charting the progress of reionization through the clustering
of Ly-$\alpha$ galaxies.\\ 
$\bullet$ Determining the source of reionization.\\ 
$\bullet$ Studying the emergence of the first galaxies over
cosmologically representative volumes.
\end{NR2}
}


\subsection{The clustering of Lyman-$\alpha$ emitters as a probe of reionization}

Cosmological simulations of the reionization process predict that the
highly-clustered, high-redshift sources of Lyman-continuum photons
will lead to an inhomogeneous distribution of ionized regions; the
reionization process is expected to proceed inside-out, starting from
the high-density peaks where the galaxies form. Thus, as demonstrated
by the state-of-the-art simulations shown in Fig.~\ref{fig:reioniz},
reionization is predicted to be highly patchy in nature. This
prediction is already gaining observational support from the latest
large-area surveys for Lyman-$\alpha$ emitters at $z \simeq 6.5$,
where it has been found that, depending on luminosity, their number
density varies by a factor of $2-10$ between different $\simeq
1/4$\,deg$^2$ fields \citep{Ouchi:10,Nakamura:11}. It is thus clear
that surveys over many square degrees are required to gain a
representative view of the Universe at z$>$6. Crucially, with such a
survey, the differential evolution and clustering of Lyman-break
galaxies and Lyman-$\alpha$ emitting galaxies can be properly measured
for the first time, offering a key signature of the reionization
process.

As has been well-demonstrated over recent years, galaxies at
high-redshift can be very effectively selected on the basis of {\it
  either} their redshifted Lyman break (the sudden drop in emission
from an otherwise blue galaxy, due to inter-galactic absorption at
wavelengths $\lambda_{rest} < 1216$\AA), or their redshifted
Lyman-$\alpha$ emission.  The former class of objects are termed
Lyman-Break Galaxies (LBGs) while the latter are termed Lyman-$\alpha$
Emitters (LAEs). In principle, LAEs are simply the subset of those
LBGs which display detectable Lyman-$\alpha$ emission, but the current
sensitivity limitations of broad-band near-infrared imaging over large
areas has meant that narrow-band imaging has been successfully used to
yield samples of lower-mass galaxies which are not usually identified
as LBGs \citep[e.g.,][]{Ono:10}. Nevertheless, as demonstrated by
spectroscopic follow-up of complete samples of bright LBGs
\citep[e.g.,][]{Stark:10,Vanzella:11,Schenker:12}, the fraction of
LBGs which are LAEs as a function of redshift, mass, and environment
is a potentially very powerful diagnostic of both the nature of the
first galaxies, and the physical process of reionization.

With the unique combination of deep, wide-area near-infrared imaging,
provided by surveys such as {\sl Euclid}, and deep, complete follow-up
near-infrared spectroscopy, made possible with \chronos, we now
propose to fully exploit the enormous potential of this approach.  The
essential idea of using \chronos to constrain reionization is as
follows: while the Lyman-$\alpha$ luminosity of LAEs is affected both
by the intrinsic galaxy properties, {\it and} by the \HI content (and
hence reionization), the luminosity of LBGs (which is measured in the
continuum) depends only on the intrinsic galaxy properties.  Thus, a
deep, wide-area, complete survey for LBGs at z$\simeq$6--12 with
accurate redshifts secured by \chronos will deliver a definitive
measurement of the evolving luminosity function and clustering of the
emerging young galaxy population, while the analysis of the follow-up
spectroscopy will enable us to determine which LBGs reside in
sufficiently large ionized bubbles for them to also be observed as
LAEs. In order to prevent strong damping wing absorption of Ly$\alpha$
photons, a galaxy must carve out a bubble of radius $R_I$
corresponding to a redshift difference with respect to the source of
$\Delta$z$>$0.01, or around 250 physical kpc at z$\approx$8.
According to the most recent reionization history predictions from
cosmological simulations, consistent with the various reionization
constraints, the \HI fraction at this redshift is around $\chi \approx
0.4-0.7$. It is easy to show that $R_I$ for a typical galaxy with a
star-formation rate of $\dot M_* = 1$\,M$_{\odot}$\,yr$^{-1}$ is of
the same order or smaller, depending on poorly established values of
the ionizing photon escape fraction. Thus, such galaxies will be only
marginally detectable in the Ly$\alpha$ line if they are isolated. In
practice, some of these galaxies will be highly clustered and
therefore will help each other in building a \HII region which is
large enough to clear the surrounding \HI and make it transparent to
Ly$\alpha$ photons.

This argument emphasizes the importance of clustering studies of LAEs,
for which \chronos is optimally designed. A key aim is to compute
in great detail the two-point correlation function of
LAEs and its redshift evolution.  For the reasons outlined above,
reionization is expected to increase the measured clustering of
emitters and the angular features of the enhancement would be
essentially impossible to attribute to anything other than
reionization. In fact, under some scenarios, the apparent clustering
of LAEs can be well in excess of the intrinsic clustering of halos in
the concordance cosmology. Observing such enhanced clustering would
confirm the prediction that the \HII regions during reionization are
large \citep{McQuinn:07}.

As required to meet our primary science goals, the \chronos
surveys will result in by far the largest and most representative
catalogues of LBGs and LAEs ever assembled at z$>$6.
Detailed predictions for the number of LBGs as extrapolated from
existing ground-based and {\it HST} imaging surveys are deferred to
the next subsection. However, here we note that the line sensitivity
of the 100 deg$^2$ spectroscopic survey will enable the identification
of LAEs with a Ly$\alpha$ luminosity $\geq 10^{42.2} \,{\rm erg \,
  s^{-1}}$, while over the smaller, ultra-deep 10 deg$^2$ survey this
line-luminosity limit will extend to $\geq 10^{41.6} \,{\rm erg \,
  s^{-1}}$. Crucially this will extend the Lyman-$\alpha$
detectability of LBG galaxies at $z \simeq 8$, with brightness $J
\simeq 27$ (AB mag), down to ``typical'' equivalent widths of $\simeq
15$\AA\ \citep{Stark:10,Vanzella:11,CLake:12,Schenker:12}.

The total number of LAEs in the combined \chronos surveys will
obviously depend on some of the key unknowns that \chronos is
designed to measure, in particular the fraction of LBGs which display
detectable Ly$\alpha$ emission as a function of redshift, mass and
environment. However, if the observed LAE fraction of bright LBGs at
$z \simeq 7$ is taken as a guide, the \chronos surveys will
uncover $\sim 10,000$ LAEs at z$>$6.5.

\subsection{The emerging galaxy population at z$>$7, and the supply 
of reionizing photons}

\chronos will provide a detailed spectroscopic characterization
of an unprecedently large sample of LBGs and LAEs.  Crucially, as well
as being assembled over representative cosmological volumes of the
Universe at z$\simeq$6--12, these samples will provide excellent sampling of the 
brighter end of the galaxy UV luminosity function at early epochs.
As demonstrated by the most recent work on the galaxy luminosity 
function at $z \simeq 7 - 9$ \citep{McLure:13}, an accurate determination of the 
faint-end slope of the luminosity function (crucial for 
understanding reionization) is in fact currently limited by 
uncertainty in $L^*$ and $\phi^*$. Consequently, 
a large, robust, spectroscopically-confirmed sample of brighter LBGs 
over this crucial epoch is required to yield definitive measurements 
of the evolving luminosity
functions of LBGs and LAEs.

Leaving aside the uncertainties in the
numbers of LAEs discussed above, we can establish a reasonable
expectation of the number of photometrically-selected LBGs which will be 
available for \chronos spectroscopic follow-up by the time 
of the mission. For example, scaling from existing {\it HST} and
ground-based studies, the `Deep' component of the {\sl Euclid} 
survey (reaching $J \simeq 26$, 5-$\sigma$
over $\simeq 40$\,deg$^2$), 
is expected to yield $\simeq 6000$ 
LBGs in the redshift range 6.5$<$z$<$7.5 with $J < 26$ (selected as ``$Z$-drops''),
$\simeq 1200$
at 7.5$<$z$<$8.5(``$Y$-drops''), and 
several hundred at z$>$8.5(``$J$-drops'')
\citep{Bouwens:10,Bowler:12,McLure:13}.

Therefore, the planned spectroscopic follow-up
over 10\,deg$^2$, will be able to target (at least)
$\simeq 1500$  
LBGs in the redshift range 6.5$<$z$<$7.5, 
$\simeq 300$ in the redshift 
bin 7.5$<$z$<$8.5, and an as yet to be determined number of 
candidate LBGs at 8.5$<$z$<$9.5.
The proposed depth and density of the \chronos near-infrared
spectroscopy will allow detection of Ly$\alpha$ line emission from these galaxies 
down to a 5-$\sigma$ flux limit 
$1 \times 10^{-18}\, {\rm  erg\, cm^{-2} s^{-1}}$, 
enabling rejection of any low-redshift interlopers, determination of the LAE
fraction down to EWs of $\simeq 10$\AA, and accurate spectroscopic redshifts 
for the LAE subset.

\subsection{The contribution of AGN to reionization \& the early growth of black holes}

SDSS has revolutionised studies of quasars at the highest redshifts,
and provided the first evidence that the epoch of reionization was
coming to an end around z$\simgt$6 \citep{Becker:01}. As with the
studies of galaxies discussed above, pushing to higher redshifts is
impossible with optical surveys, regardless of depth, due to the fact
that the Gunn-Peterson trough occupies all optical bands at z$>$6.5.
Therefore, to push these studies further in redshift needs deep
wide-field surveys in the near-infrared.

The wide-area, ground-based VISTA near-infrared public surveys such as VIKING 
and the VISTA hemisphere survey are slowly beginning to uncover a few bright quasars
at z$\simeq$7 \citep[e.g.,][]{Mortlock:11}, and it is to be expected that {\it Euclid} 
will be able to provide a good determination of the very bright end of the QSO luminosity 
function at z$>$6. However, the shape of the QSO luminosity function at these 
redshifts can only be studied with 
detailed near-infrared spectroscopy 
over a significant survey area. This is the only direct way to
properly determine the contribution of accreting black holes to the
reionization of the Universe and constrain the density of black-holes
within the first Gyr after the Big Bang; \chronos's combination
of depth and area provides the ideal way in which to measure the
evolving luminosity function of quasars at $ 6.5 < z < 10$.

\end{multicols}

\begin{NR1}[userdefinedwidth=17.5cm]
\section{Additional science cases}
\label{sec:Add}
\end{NR1}

\begin{figure*}
\begin{minipage}[h]{16cm}
  \centering
   \includegraphics[width=43mm]{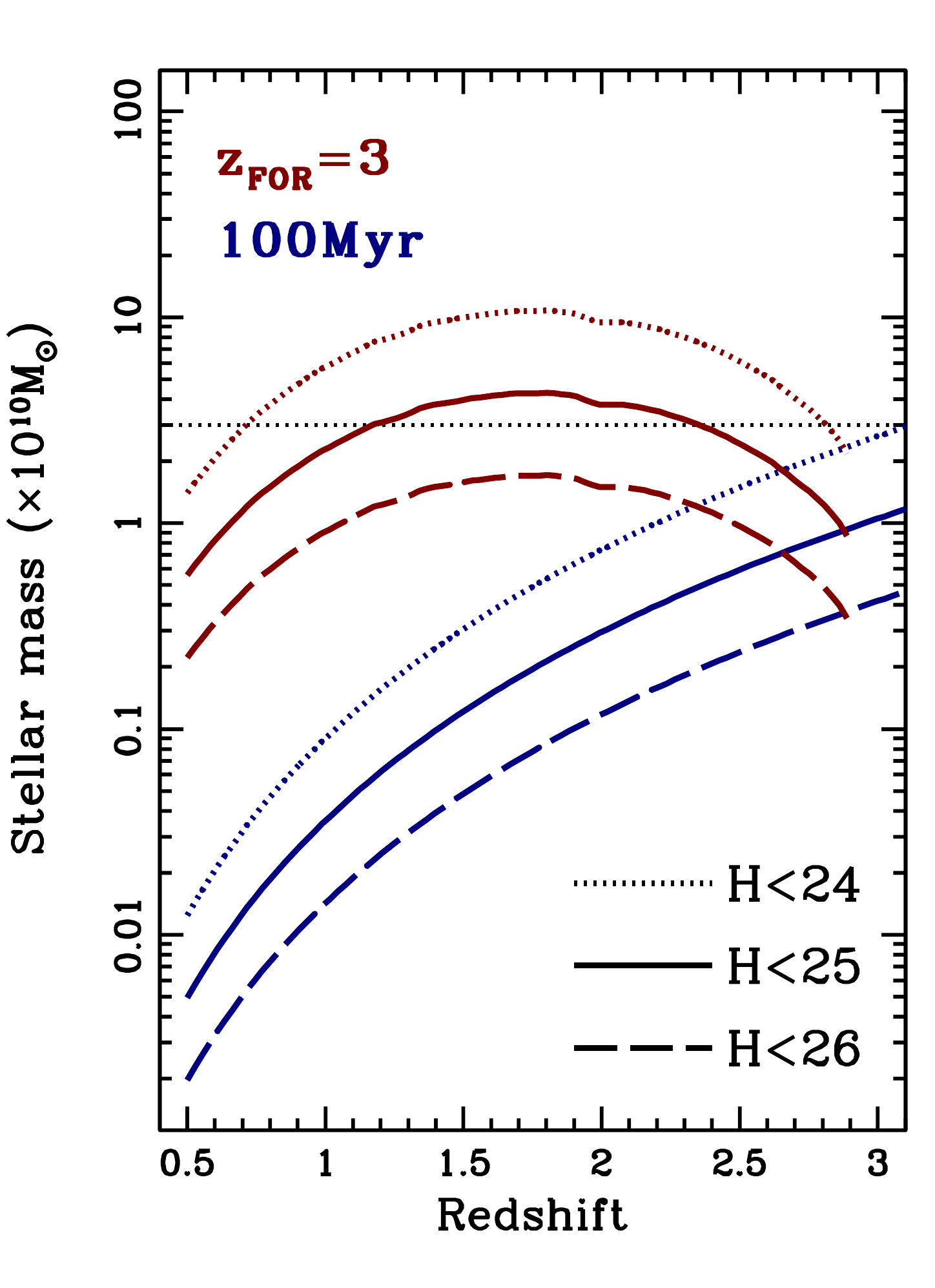}
   \includegraphics[width=43mm]{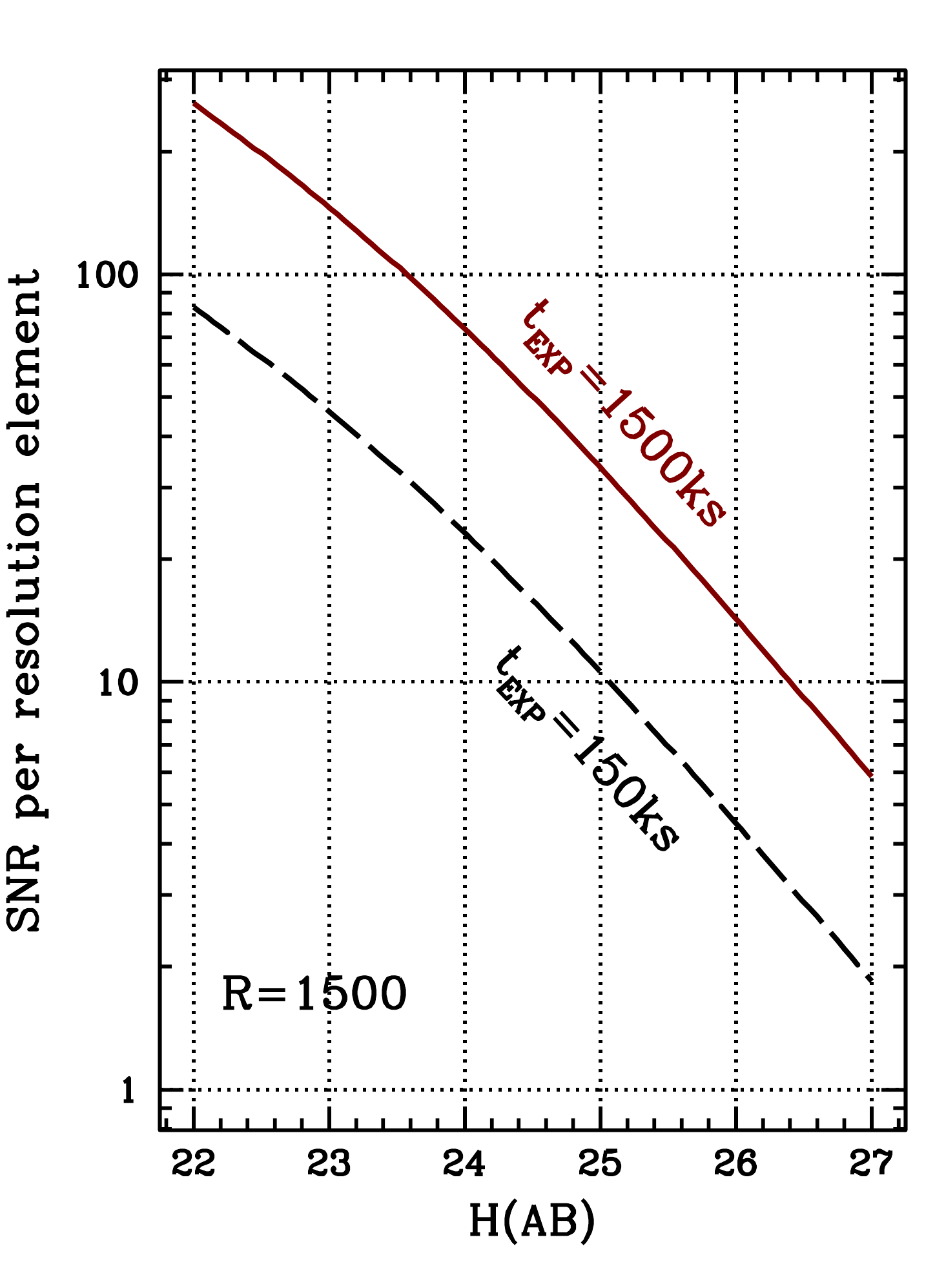}
   \includegraphics[width=43mm]{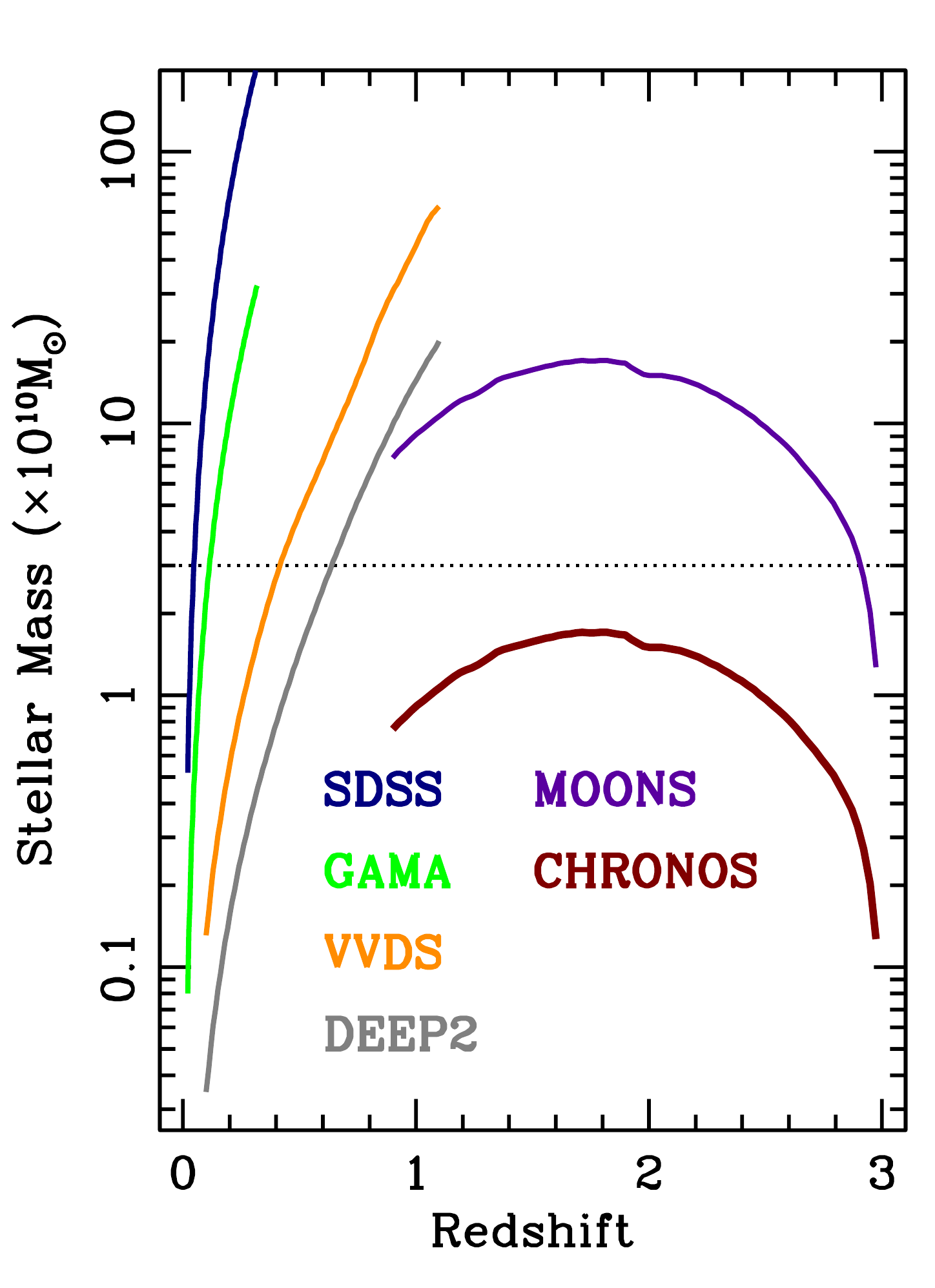}
\end{minipage}
\caption{\small {\bf Left:} Stellar mass versus redshift for three
  limiting-magnitude surveys, as labelled. The red lines correspond to
  old stellar populations, whereras the blue lines trace a more
  luminous, young population, corresponding to the typical age of a
  star forming galaxy. {\bf Middle:} SNR per resolution element at
  $R\equiv\lambda/\Delta\lambda=1500$ for a 2.5m (diameter) collecting
  area, 20\% total efficiency, a zodiacal light (minimal) level
  of H$_{\rm AB}$=25 and typical readout noise and dark current of
  cutting-edge NIR arrays (RN=4e, 0.01 e/s, respectively).  We note
  that the SNR is estimated in the {\sl continuum}, a much more
  stringent requirement than emission line estimates, typical of
  cosmology-orientated surveys. {\bf Right:} Comparison of \chronos\
  (wide survey) with a range of optical and NIR spectroscopic surveys
  with similar spectral resolution. The curves track the lower limit
  in stellar mass for an old population (i.e. the worst case
  scenario).}
\label{fig:Hlimit}
\vspace{-0.3cm}
\end{figure*}

\begin{multicols}{2} 
Although this white paper deals with the core science driver
of galaxy formation and evolution in the spectral window
$1\simgt$z$\simgt 12$, the legacy side of \chronos is
immense, and additional science projects can be addressed.
Among them, we list a few relevant cases below:
\begin{compactitem}
\item[i)]{Transients: the planned 5-year mission can
  accommodate the spectroscopic follow up of transients, most notably
  high redshift supernov\ae, allowing us not only to confirm the type
  of supernova, but spectroscopic features could be used to understand
  their properties and evolution, relevant to precision cosmology
  studies \citep{FK:11}.}
\item[ii)]{Cosmology: \chronos will enable cosmological model
  testing beyond {\sl Euclid}. As an example, measurements of the
  velocity field, galaxy bias, and lensing potential simultaneously
  will enable a measurement of general single scalar-field models
  \citep[e.g.,][]{Amen:12}. The deep redshift range would also
  constrain early-dark energy models, complementing the {\sl Euclid}
  cosmology objectives using techniques such as those used by
  \citet{Mandel:12} in SDSS.}
\item[iii)]{Brown dwarves: The \chronos survey could include a
  programme to explore cool T-dwarves out to a few hundred pc,
  allowing us to determine the scale height of this population. By
  targeting nearby star forming regions, we can probe
  the IMF down to Jupiter-size masses.}
\end{compactitem}

\end{multicols}

\begin{NR1}[userdefinedwidth=17.5cm]
\section{Survey requirements}
\label{sec:survey}
\end{NR1}

\begin{multicols}{2} 

The core requirement of the survey is the apparent magnitude limit to
obtain a complete sample selected in stellar mass, with acceptable SNR
per resolution element for the study of the underlying stellar
populations. Any other criteria commonly approached in surveys
(e.g. selection in luminosity or colour) will bias the sample. We use
stellar population synthesis models \citep{BC:03} to estimate the
apparent magnitude with respect to stellar mass and redshift for two
extreme scenarios (Fig.~\ref{fig:Hlimit}, left panel).  An old
population (red lines) will lack the most massive and luminous stars,
therefore appearing significantly fainter than a younger population
with the same mass (blue lines). The age range used in the figure
covers a conservative interval as obtained, e.g., from spectroscopic
observations of the local Universe \citep{Gallazzi:05}. The figure
shows that at H$_{\rm AB}\sim$26, we will obtain a complete sample out to
z$\sim$3 for galaxies with stellar mass around $1-2\times
10^{10}$\,M$_\odot$. This is a {\sl conservative} estimate, as
observations suggest that there is a strong trend towards younger ages
in low mass galaxies, with the characteristic star-forming galaxy
appearing more massive at higher redshift \citep{PPG:08}. This trend
implies that the majority of low-mass galaxies will be younger,
allowing us to reach a completeness level -- if we relax the
constraint regarding old populations -- at lower stellar masses,
possibly around $10^9$\,M$_\odot$, covering an unprecedented range of
galaxy mass over the z=1$-$3 redshift window. The middle panel of
Fig.~\ref{fig:Hlimit} gives an estimate of the SNR per resolution
element (at R=1500) achieved for two exposure times, as labelled. We
envisage a two-tiered survey, comprising a wide-deep survey covering
100\,deg$^2$ (corresponding to the dashed black line) and an
ultra-deep survey with a 10 times longer integration over 10\,deg$^2$
(solid red line). We emphasize here that a proper characterization of the
stellar populations from spectroscopic data requires SNR$\simgt
5-10$. This figure assumes a low zodiacal and thermal background
level, with a spatial resolution of 0.3\,arcsec, and typical detector
noise for the type of available arrays (e.g., Teledyne Hawaii
4RG). The righmost panel of Fig.~\ref{fig:Hlimit} compares the ability
of \chronos to obtain a mass-limited sample out to a chosen
redshift with recent or planned spectroscopic surveys at similar
resolution. We note that neither {\sl Euclid} nor {\sl WFIRST} 
estimates are
included in the figure, as their low-resolution, slitless grism
spectra are not capable of achieving the goals of this white
paper\footnote{Nevertheless, as a reference, the {\sl Euclid}
Definition Study Report states that at z$\geq$1.5 only galaxies with 
a stellar mass $>4\times 10^{11}$M$_\odot$ will provide useable
spectra for the analysis of the populations \citep{EuclidRB}.}.
{\sl MOONS} is clearly the best option at present in the
z=1$-$3 redshift range, however, the signal in the continuum will be
weak unless very young populations are considered. Only \chronos
can provide the collecting power and wide field of view to tackle in
an unbiased way the analysis of galaxies at the peak of activity.

As regards to the required total areal coverage on the sky, we use as
reference the SDSS, whose high quality spectroscopic data can be
extended out to, at most, z$\simlt$0.2, covering a comoving volume of
$5.5\times 10^{-5}$\,Gpc$^3$ per square degree. Over the proposed
z$\sim$1--3 range, we have 0.02\,Gpc$^3$ per deg$^2$. Hence, in order
to probe the environment in detail comparable to the $\sim
10^4$\,deg$^2$ of SDSS/DR7 \citep{DR7}, we need around 30\,deg$^2$.
In addition, we expect environment to evolve significantly between the
SDSS baseline and the goal of \chronos. We use a large
cosmological simulation \citep[Millennium,][]{Springel:05} to find an
evolution in the number of a factor of $\sim$2--3 for groups with halo
mass M$_{\rm halo}\simgt 10^{12}$\,M$_\odot$ at z=0. Therefore, the
general survey should target around 100\,deg$^2$, putting this
project outside of the reach of {\sl JWST} or any of the extremely 
large telescopes on the ground. SDSS has also shown
that datasets comprising $\sim 1$ million spectra are necessary to split the
sample with respect to the many properties under consideration (velocity
dispersion, luminosity, mass, environment, etc). Finally, an extrapolation
of the \citet{Muzz:13} data using a fit to a Schechter law gives
a number density of $1.2\times 10^{5}$ galaxies per square degree at the
H$_{\rm AB}$=26 level in the z$\sim$1--3 range.

\myfigure{
\begin{NR2}[userdefinedwidth=85mm]
\begin{center}
\vspace{-5mm}
... in a nutshell
\end{center}
$\bullet$ At H$_{\rm AB}$=26 we expect $\simgt$100,000 galaxies per square
degree at z$\sim$1--3\\ 
$\bullet$ Completeness down to a stellar mass 
$\sim 10^{10}$M$_\odot$ (z$\simlt$3) {\sl for any population}.\\
$\bullet$ Two surveys: 100\,deg$^2$ and
10\,deg$^2$ extending over the z$\simgt$1 environments probed by SDSS 
at z$\simlt$0.1.\\
$\bullet$ The final dataset will comprise $\sim 1-2$ million high-quality spectra.
\end{NR2}
}

\end{multicols}

\begin{NR1}[userdefinedwidth=17.5cm]
\section{Strawman Mission Concept}
\end{NR1}

\begin{multicols}{2} 

\subsection{Mission Profile}
\label{sec:mission}

As an infrared survey mission, the preferred orbit for \chronos
is at the low background L2 point, with heritage from {\it Herschel}
and {\it Planck} operations and, in the future, from {\it Gaia}, 
{\sl JWST} and {\it Euclid}.
An Ariane 5 ECA or ME launcher provides excellent payload margin with
a limit to L2 in excess of 6.2\,tonnes. The standard fairing has a
length of 12.7\,m and a diameter of 4.6\,m which can easily accommodate
the proposed \chronos spacecraft configuration; these dimensions
would allow a full 2.5m diameter f/1.2 telescope to be deployed
without any dynamic mechanisms.  Other launcher options could be
considered, depending on launch date.  Once L2 has been reached, the
$\Delta$v requirements for orbit station and formation keeping are
small ($< 75$~m~s$^{-1}$~year$^{-1}$).

\begin{figure*}
\RawFloats\CenterFloatBoxes
\begin{floatrow}[2]
\ffigbox[\FBwidth]{%
   \includegraphics[width=78mm]{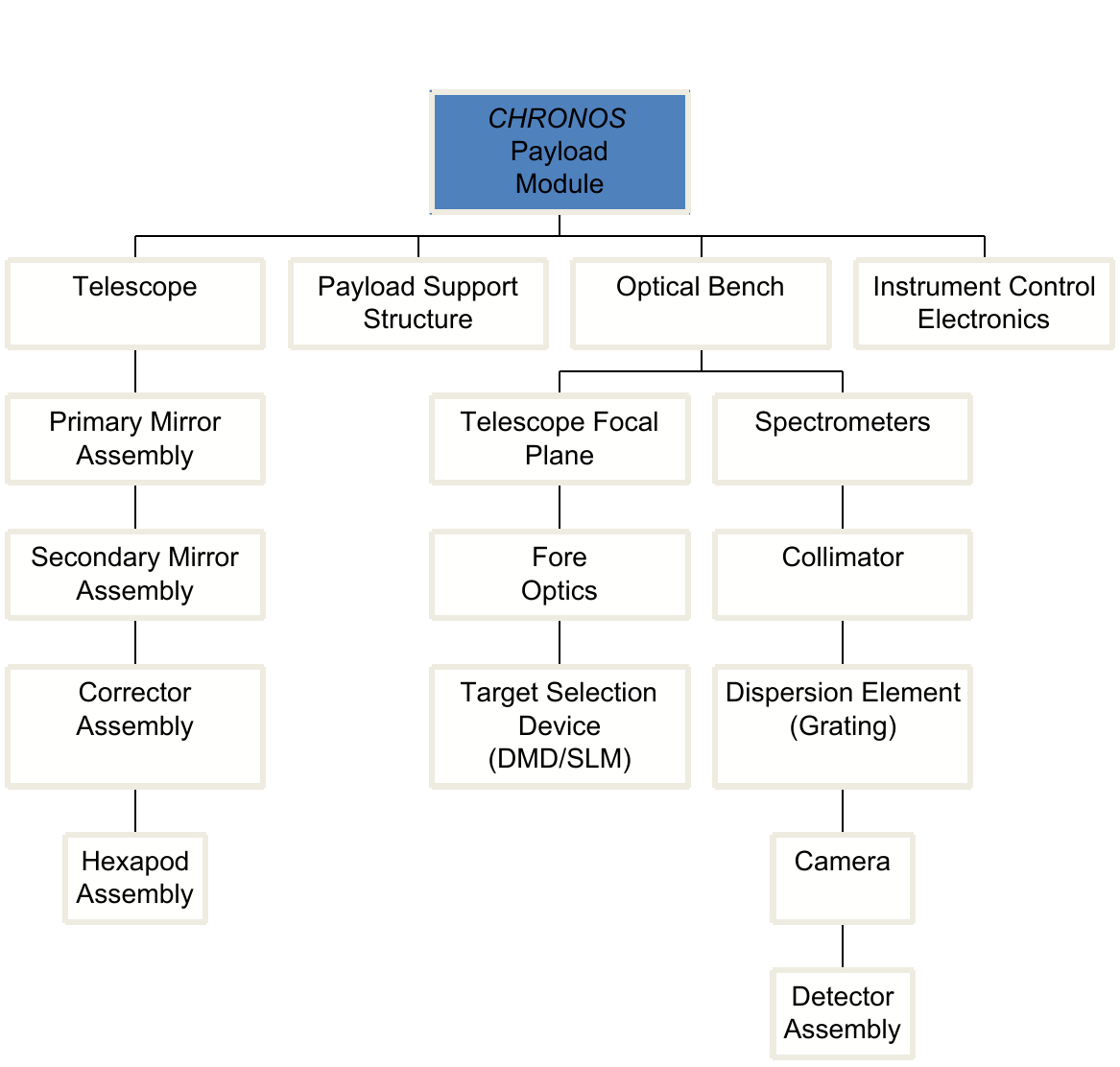}
\caption{\small Subsystem breakdown of the payload module hardware.}
\label{fig:payload}}{
}
\hspace{8mm}
\capbtabbox{
\begin{tabular}{|l|l|}
\hline
{\bf Item} & {\bf Payload Requirements} \\
\hline
\multicolumn{2}{|l|}{{\bf Telescope}} \\
\hline
Primary mirror & 2.5 metre diameter \\
\hline
Field of View & 1.0 degree diameter \\
\hline
Image quality & EE(80)$ < 0.3$\,arcsec \\
\hline
\multicolumn{2}{|l|}{{\bf Spectrometers}} \\
\hline
Target multiplex & $\sim5000$ objects per pointing \\
\hline
Field of view & 0.2 deg$^2$ (total, 8 spectrometers) \\    
\hline
Spectral coverage & $0.9\mu$m to $1.8\mu$m \\
\hline
Spectral resolution & R~$\sim 1500$ \\
\hline
Throughput & $>20$\% including detectors\\
\hline
\multicolumn{2}{|l|}{{\bf Spacecraft}} \\
\hline
Mass & $<4000$ kg \\
\hline
Volume & \\
Diameter & 3500 mm \\
Length &  7000 mm \\
\hline
\end{tabular}
}{%
\caption{\small Summary of the key performance requirements of the \chronos payload.}
\label{tab:payload}
}
\end{floatrow}
\vspace{-0.3cm}
\end{figure*}

The \chronos survey strategy will follow that adopted for the
{\it Euclid} deep-field programme, with frequent revisits to the same
field centres to build up S/N on faint targets, ameliorate the
contamination effects in crowded fields, and provide useful cadence
for serendipitous studies of high-redshift supernovae and gamma-ray
bursts. The regular layout of the proposed \chronos focal plane
will allow a simple tiling strategy to cover contiguous areas of the
100 deg$^2$ (deep) and 10 deg$^2$ (ultra-deep) fields. For a 5 year  
mission, at a 70\% operational efficiency, the draft survey plan calls
for visiting one pointing per week (deep) or one pointing per 10 weeks
(ultra-deep with 10x longer exposure). Assuming that \chronos
targets 30\% of the available sample at a multiplex of $\sim5000$, will
require $\sim 8$ mask configurations per pointing, giving a total
exposure time of 150\,ksec for the deep survey and 1500\,ksec for the
ultra-deep survey. The final galaxy samples would thus comprise
$\sim$1.5 million ($\sim$150 thousand) high quality spectra in the
deep (ultra-deep) surveys.

As a dedicated survey mission, the ground segment can be kept
relatively simple. Target definition for the survey will come from the
optical-infrared imaging in the {\it Euclid} deep fields or, as a
fallback, from {\it LSST} and {\it VISTA} ground-based deep survey
(AB$\sim 27.5$ optical and AB$\sim 24.5$ infrared respectively). Fast
data analysis will be required only for the transient detection
programme. The downstream data rate will be approximately 50 GB/day
after compression (depending on the number of intermediate detector
samples are transmitted); assuming a typical K-band rate of transfer
to the ground of 50 Mbit/s, all of the data can be transferred to the
ground with a contact time of 3 hours per 24 hours.

\subsection{Payload Description}
\label{sec:payload}

The science requirements for \chronos drive the choice of a
telescope with a 2.5m aperture and a 1\,deg field-of-view
feeding eight identical multi-object slit-based spectrometers with
moderate spectral resolution and good background
subtraction. Selection of the science targets for the spectrometers
can be achieved by using a digital micromirror device (DMD) or other
form of spatial light modulator. There is no scientific need to make
\chronos into a multi-purpose observatory so the spectrometers
have a single fixed resolving power (R~$\sim 1500$).  The DMD or
spatial light modulator can be used to shut off the signal to the
detectors, so no mechanisms are required.
The payload module hardware breaks down into testable sub-systems as
shown in Figure \ref{fig:payload}. The key performance parameters of
the \chronos payload are shown in Table \ref{tab:payload}.

\subsubsection{Telescope Assembly}
\label{subsec:telescope}

The 2.5-metre telescope could be a Korsch or Ritchey-Chr\'etien design
for which a three-element field corrector which would give a 
field-of-view of 1\,degree at f/3 to feed the eight spectrometers. Assuming a
Ritchey-Chr\'etien design, this could be 
optimised to provide uniform image quality across the whole field with
an image quality (EE80$\sim$0.3\,arcsec) which is matched to the
intrinsic size of galaxies at high redshift. Both the primary (f/1.2)
and secondary (f/3) mirrors are hyperboloids. The secondary mirror
(M2) is 0.9m in diameter and its mounting incorporates light sources
for calibration of the spectrometer.  While SiC will provide excellent
performance if a 2.5m optical-quality mirror can be fabricated,
lightweighted Zerodur would also be possible.  The baseline design
assumes that both M1 and M2 are fabricated from lightweighted Zerodur
and are supported directly from the telescope support structure. A
central light baffle incorporates the three fused silica corrector
elements in the R-C design (two aspheric surfaces).

The telescope support structure is a SiC or CFRP space frame which
supports all the hardware of the spacecraft, and under which is
mounted the Instrument Optical Bench and the Instrument Service
Module. The upper section of the telescope structure consists of a
triangular frame, the corners of which act as structural nodes for the
M1 backing structure, the M2 hexapod and the instrument optical
bench. The total length of the telescope is approximately 5\,metres
with a mass of 600\,kg (excluding payload and service module).

\begin{figure*}
\RawFloats\CenterFloatBoxes
\begin{floatrow}[3]
\ffigbox[\FBwidth]{%
   \includegraphics[width=60mm]{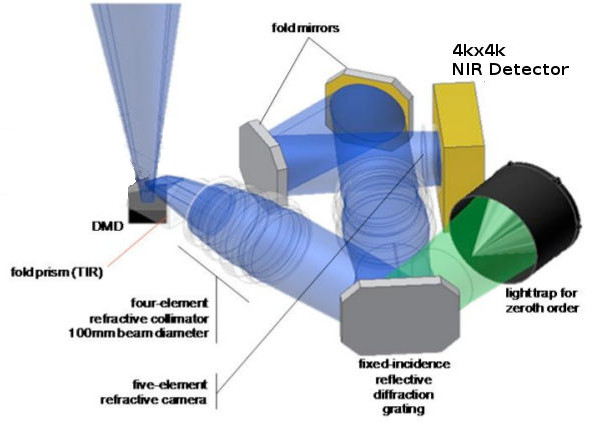}
\caption{\small Close-up of a single spectrometer
  channel.}
\label{fig:spec}}{
}
\hspace{2mm}
\ffigbox[\FBwidth]{%
   \includegraphics[width=55mm]{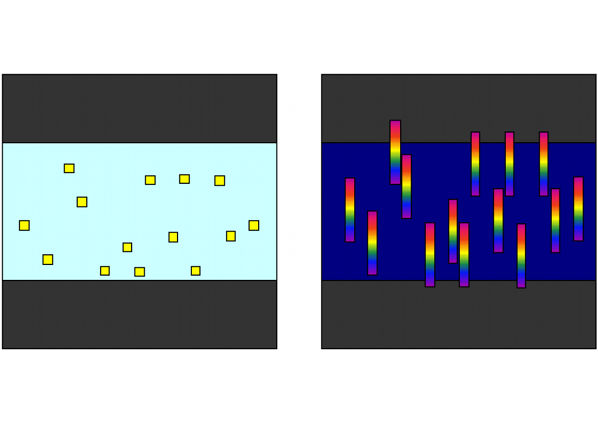}
\caption{\small Schematic of multi object target selection using DMDs.}
\label{fig:DMD}}{
}
\hspace{2mm}
\capbtabbox{
\vspace{8mm}
\begin{tabular}{|l|r|}
\hline
{\bf Phase A} & 30~Meuro\\
\hline
{\bf Phase B} & 200~Meuro\\
\hline
{\bf Phase C/D} & 600~Meuro\\
\hline
{\bf Launch} & 200~Meuro\\
\hline
{\bf Phase E/F} & 200~Meuro\\
\hline\hline
{\bf Total} & 1230~Meuro\\
\hline
\end{tabular}
}{%
\vspace{4mm}
\caption{\chronos\ lifecycle cost estimate, including national hardware contributions.}
\label{tab:cost}
}
\end{floatrow}
\vspace{-0.3cm}
\end{figure*}

\subsubsection{Spectrometer}
\label{subsec:spectrometer}

At the heart of the \chronos concept are a set of eight 
multi-object spectrometers (Fig.~\ref{fig:spec}), each 
capable of delivering complete samples of moderate resolution (R~$\sim$1500) 
near-infrared spectra for high redshift galaxies down to a magnitude
limit of H$_{\rm AB}\sim$26\,mag. To reach this faint limit requires
'multi-slit' spectroscopy with a target selection mechanism which is
compatible with space operations.  
Our baseline approach is to use the Texas Instruments (TI) digital
micro mirror devices (DMDs) which were originally proposed for the
SPACE mission \citep{CI:09}. These are available in formats up to
$2048\times1080$ pixels with a pitch of 13.68 $\mu$m and are currently
at a technology readiness level of TRL~$\sim 4$ \citep{ZA:10}. Each
of the individual micromirrors on the DMD can be switched into an 'ON'
or 'OFF' position to define a {\it virtual} slit of $1.2 \times 0.4$
arcsec, centred on the target of interest, thus replicating the
multislit masks used in ground-based spectroscopy of faint targets
(Figure~\ref{fig:DMD}).

Simulations of the targeting efficiency of the DMD at the magnitude
limit of the survey, indicate that each spectrometer can obtain the
spectra for $\sim 600$ targets simultaneously, without spectral
overlaps, giving a multiplex of $\sim 4800$ targets with eight
spectrometers covering a total field of $\sim 0.2$\,deg$^{2}$.

To mitigate against qualification and availability of DMD devices in
the timescale of an L2/L3 mission, a parallel technology development
study should also be initiated early in the project to assess the
technology readiness of other forms of target selection devices,
including liquid crystal spatial light modulators and pupil beam
steering devices.

The entrance apertures of the spectrometers will be positioned
symmetrically within the telescope field of view to allow a simple
step-and-stare operation to tile the sky contiguously. 
Each spectrograph will use refractive collimators and cameras,
feeding a 4kx4k HgCdTe infrared array. A 60 lines/mm
grating is used to produce a Nyquist sampled spectrum covering the
range 0.9-1.8 $\mu$m.

\subsection{Operational Model}
\label{sec:operations}

The \chronos operational model follows the usual lines of a
survey-type project. The satellite will operate autonomously except
for defined ground contact periods during which housekeeping and
science telemetry will be downlinked, and the commands needed to
control spacecraft and payload will be uploaded. The data rate is
around 50\,GB/day which is easily handled with current data-processing  
systems. A data model for the mission will be developed in
collaboration with ESA. Based on the data model, an archive system
will be built, enabling data archiving, data processing and
distribution of all \chronos observations with appropriate levels
of processing, including all the necessary ancillary information.

\subsection{Programmatics and Cost}
\label{sec:cost}

\chronos is envisaged as a typical science mission with ESA
having overall control, but with a major contribution from a
consortium of European institutes in the form of the science payload
and ground segment. The mission has been designed to ensure that
technologies with space heritage or high TRL are used where
possible. The DMDs are an exception to this, although good progress
has already been made in developing these for space application for
the original {\it SPACE/Euclid} mission concepts. We believe that this
technology can be further developed in good time for the \chronos
mission and will thus not be a schedule driver. An approximate
cost per phase is shown in Table \ref{tab:cost}.

\myfigure{
\begin{NR2}[userdefinedwidth=85mm]
\begin{center}
\vspace{-5mm}
... in a nutshell
\end{center}
$\bullet$ 2.5m telescope, Korsch or Ritchey-Chr\'etien in SiC\\
$\bullet$ 8 spectrometers, R=1500, 4800 multiplex.\\
$\bullet$ 5 year mission at L2.\\
$\bullet$ Ariane 5 launcher.
\end{NR2}
}

\clearpage


\setlength{\bibsep}{3.0pt}

{\fontsize{9pt}{9pt}\selectfont

}


\end{multicols}

\end{document}